\documentclass[twocolumn,tighten]{aastex62}
\usepackage{graphicx,amsmath,natbib}

\shorttitle{Type Ia Supernovae in Dwarf Galaxies}
\shortauthors{Kirby et al.}
\newcommand{\citeposs}[1]{\citeauthor{#1}'s (\citeyear{#1})}

\newcommand{\sifeiaulimscl}{-0.86}

\newcommand{\nifeiamindsph}{Ursa Minor}
\newcommand{\nifeiamin}{-0.68}
\newcommand{\nifeiaminerrlo}{0.09}
\newcommand{\nifeiaminerrhi}{0.07}
\newcommand{\nifeiamaxdsph}{Leo~II}
\newcommand{\nifeiamax}{-0.13}
\newcommand{\nifeiamaxerrlo}{0.04}
\newcommand{\nifeiamaxerrhi}{0.03}

\newcommand{\ndsph}{1819}

\begin{document}

\title{Evidence for Sub-Chandrasekhar Type~Ia Supernovae from Stellar
  Abundances in Dwarf Galaxies\footnote{The data presented herein were
    obtained at the W.~M.~Keck Observatory, which is operated as a
    scientific partnership among the California Institute of
    Technology, the University of California and the National
    Aeronautics and Space Administration. The Observatory was made
    possible by the generous financial support of the W.~M.~Keck
    Foundation.}}

\correspondingauthor{Evan N. Kirby}
\email{enk@astro.caltech.edu}

\author[0000-0001-6196-5162]{Evan N. Kirby}
\affiliation{California Institute of Technology, 1200 E.\ California Blvd., MC 249-17, Pasadena, CA 91125, USA}

\author{Justin L. Xie}
\affiliation{Harvard College, 28 Fernald Dr., Cambridge, MA 02138, USA}

\author{Rachel Guo}
\affiliation{Harvard College, 28 Fernald Dr., Cambridge, MA 02138, USA}

\author[0000-0002-4739-046X]{Mithi A.~C. de los Reyes}
\affiliation{California Institute of Technology, 1200 E.\ California Blvd., MC 249-17, Pasadena, CA 91125, USA}

\author[0000-0002-9908-5571]{Maria Bergemann}
\affiliation{Max-Planck Institute for Astronomy, D-69117, Heidelberg, Germany}

\author{Mikhail Kovalev}
\affiliation{Max-Planck Institute for Astronomy, D-69117, Heidelberg, Germany}

\author[0000-0002-9632-6106]{Ken J. Shen}
\affiliation{Department of Astronomy and Theoretical Astrophysics Center, University of California, Berkeley, CA 94720, USA}

\author{Anthony L. Piro}
\affiliation{The Observatories of the Carnegie Institution for Science, 813 Santa Barbara St., Pasadena, CA 91101, USA}

\author{Andrew McWilliam}
\affiliation{The Observatories of the Carnegie Institution for Science, 813 Santa Barbara St., Pasadena, CA 91101, USA}


\begin{abstract}

There is no consensus on the progenitors of Type~Ia supernovae
(SNe~Ia) despite their importance for cosmology and chemical
evolution.  We address this question by using our previously published
catalogs of Mg, Si, Ca, Cr, Fe, Co, and Ni abundances in dwarf galaxy
satellites of the Milky Way (MW) to constrain the mass at which the
white dwarf (WD) explodes during a typical SN~Ia\@.  We fit a simple
bi-linear model to the evolution of [X/Fe] with [Fe/H], where X
represents each of the elements mentioned above.  We use the evolution
of [Mg/Fe] coupled with theoretical supernova yields to isolate what
fraction of the elements originated in SNe~Ia.  Then, we infer the
[X/Fe] yield of SNe~Ia for all of the elements except Mg.  We compare
these observationally inferred yields to recent theoretical
predictions for two classes of Chandrasekhar-mass ($M_{\rm Ch}$) SN~Ia
as well as sub-$M_{\rm Ch}$ SNe~Ia.  Most of the inferred SN~Ia yields
are consistent with all of the theoretical models, but [Ni/Fe] is
consistent only with sub-$M_{\rm Ch}$ models.  We conclude that the
dominant type of SN~Ia in ancient dwarf galaxies is the explosion of a
sub-$M_{\rm Ch}$ WD\@.  The MW and dwarf galaxies with extended star
formation histories have higher [Ni/Fe] abundances, which could
indicate that the dominant class of SN~Ia is different for galaxies
where star formation lasted for at least several Gyr.

\end{abstract}



\section{Introduction}
\label{sec:intro}

Type~Ia supernovae (SNe~Ia) are some of the most important events in
astrophysics.\footnote{See reviews by \citet{yun00}, \citet{hil13},
  \citet{mao14}, \citet{mag17}, and \citet{sei17}.}  They are the
basis for the Nobel Prize-winning measurement of cosmological
acceleration \citep{rie98,per99}.  They are also the origin of most of
the iron and some of the elements adjacent in atomic number to iron
(Fe-peak elements) in the Galaxy \citep{nom84a}.  Despite their
importance, there remain outstanding questions about the physics of
the explosions and the nature of their progenitors.  They result from
the explosions of carbon/oxygen white dwarfs \citep[WDs;][]{arn69} in
binary systems, but the details of the burning and even the numbers of
WDs involved are hotly debated.  Resolving these questions might
provide a physical basis for the use of SNe~Ia as standardizable
candles \citep{phi93}, thus justifying their widespread use as
cosmological tools.

Models of SNe~Ia have been developed over the past five decades.  The
first models supposed that a WD grew in mass by the accretion of
hydrogen from a red giant companion \citep{whe73}.  When the WD neared
the Chandrasekhar mass ($M_{\rm Ch}$), its core reached sufficiently
high temperature and density to ignite carbon.

\citet{nom84b} first calculated the nucleosynthesis of SNe~Ia exploding
with a mass near $M_{\rm Ch}$.  They found that the rate of carbon
burning strongly affected the elements produced.  If the WD {\it
  detonated} supersonically, i.e., on a timescale faster than the
dynamical time, it produced Fe-group elements but almost no Si.  In
order to explain the Si present in the spectra of SNe~Ia,
\citeauthor{nom84b}\ invoked {\it deflagration}, whereby a
carbon-burning flame consumes the WD subsonically, i.e., more slowly
than the dynamical time.

A hybrid of deflagration and detonation \citep{kho91,iwa99} can
achieve the appropriate balance of Si-group and Fe-group elements.  In
this model, an initial deflagration gives way to a detonation.  The
model is called ``delayed detonation'' or ``deflagration-to-detonation
transition'' (DDT)\@.  The deflagration burns material at high
density, but it also allows the WD to expand and become less dense
before the detonation consumes the remainder of the carbon.  The
details of the nucleosynthesis depend heavily on the structure of the
progenitor WD and the physics of the carbon ignition
\citep[e.g.,][]{leu18,byr19}, as well as the implementation of the
hydrodynamics and the dimensionality of the simulation code
\citep[e.g.,][]{mae10,sei13a}.

The near-$M_{\rm Ch}$ model has confronted some obstacles.  For
example, there are very few observed WDs with masses approaching
$M_{\rm Ch}$ \citep{gia12,kle13}, which would be the progenitors for
near-$M_{\rm Ch}$ SNe~Ia.\footnote{However, at least one candidate
  progenitor for a future near-$M_{\rm Ch}$ SN~Ia has been discovered
  \citep{tan14}.}  Stable accretion to build up a WD to $M_{\rm Ch}$
imposes very fine requirements on the accretion rate \citep{she07}.
Furthermore, a sufficient number of WDs accreting to produce the SN~Ia
rate is disfavored by X-ray observations of nearby early-type galaxies
\citep{gil10}.  Finally, there is a lack of companion stars associated
with SN~Ia remnants as would be expected in this scenario
\citep[e.g.,][]{sch12}.  One possible solution to these problems is
the detonation of a sub-$M_{\rm Ch}$ WD\@.  Sub-$M_{\rm Ch}$ WDs are
numerous enough to explain SN~Ia rates, and their range of masses can
explain the range of SN~Ia luminosities \citep{she17}.

In addition to the mass of the exploding WD, another outstanding
question is the evolutionary path that leads to the explosion.  The
progenitor systems are generally grouped into ``single-degenerate''
and ``double-degenerate'' binaries involving one or more CO WDs.  Both
types are possible for near-$M_{\rm Ch}$ and sub-$M_{\rm Ch}$ models.
The types of explosion can be classified as follows:

\begin{enumerate}
\item The original single-degenerate, near-$M_{\rm Ch}$ model invoked
  accretion of hydrogen, usually from a red giant, onto a CO WD until
  it reached $M_{\rm Ch}$ \citep[e.g.,][]{whe73}.  More recent models
  \citep{yoo03,bro16} examined the transfer of helium rather than
  hydrogen.  In this case, the He burns into C and O until the WD
  nears $M_{\rm Ch}$. \vspace{-2mm}
\item The double-degenerate, $M_{\rm Ch}$ model supposes that two WDs
  that are individually below $M_{\rm Ch}$ merge slowly \citep{web84}.
  If the merged remnant exceeds $M_{\rm Ch}$, it will explode.  This
  channel is now disfavored because it is expected to result in
  accretion-induced collapse into a neutron star
  \citep{sai85,sai04,tim94,she12}. \vspace{-2mm}
\item The single-degenerate, sub-$M_{\rm Ch}$ model is similar to case
  1.  The difference is that sub-$M_{\rm Ch}$ WDs can explode if
  they accrete He slowly enough to accumulate a critical amount before
  it ignites.  The eventual He ignition could be strong enough to
  shock the inner CO WD to thermonuclear densities and temperatures
  \citep{nom82,woo86}. \vspace{-2mm}
\item Double-degenerate, sub-$M_{\rm Ch}$ explosions are similar to
  case 2.  The difference is in the end phase of the merger.  If He
  is transferred from the surface of one WD to the other, it could
  ignite, leading to a detonation \citep{gui10,pak13,she14,tow19},
  similar to case 3.  In this case, the secondary WD would survive
  and fly away at its final orbital velocity of thousands of
  km~s$^{-1}$ \citep{she18b}.  Alternatively, a violent merger could
  directly ignite the carbon without the need for He ignition
  \citep{pak12}.
\end{enumerate}
  
\noindent
In a variation on case 2, \citet{van10} proposed that two WDs can
merge smoothly, but compressional heating from the resulting accretion
disk can ignite the carbon, even if the final mass does not exceed
$M_{\rm Ch}$.

There is also a separate case of failed explosions in which the WD
does not completely disrupt.  The idea of a failed explosion is
motivated by the existence of Type~Iax SNe, a class of sub-luminous SN
\citep{fol13}.  \citet{kro15} conjectured that Type~Iax SNe are weak
deflagrations that do not disrupt the WD and that leave behind a
remnant.  In this scenario, the partially exploded WD would become
unbound from its companion.  Thus, some hypervelocity WDs might be
associated with Type~Iax SNe \citep{ven17,rad18a,rad18b,rad19} rather
than the sub-$M_{\rm Ch}$, double-degenerate scenario \citep[case 4
  above;][]{she18b}.  Type~Iax SNe likely have a limited effect on
galactic chemical evolution because they probably do not eject much
mass.

Various individual SNe~Ia support nearly all classes of explosion.
For example, ultraviolet pulses have been interpreted as the
interaction of the SN shock with a red giant companion in SN~2012cg
\citep[][rebutted by \citealt{sha18}]{marion16}, SN~2017cbv
\citep{hos17}, and iPTF15atg \citep{cao15}, which was a peculiar
explosion similar to SN~2002es \citep{gan12,whi15}.  The first two
discoveries support the single-degenerate, near-$M_{\rm Ch}$ model.
However, H$\alpha$ emission is expected in late-time spectroscopy of
SNe~Ia arising from a H-rich donor.  With a few exceptions
\citep[i.a.,][]{ham03,kol19}, H$\alpha$ is rarely found in late-time
nebular spectra of SNe~Ia \citep{mag16,sha18,tuc19}.  Furthermore,
\citet{li11} and \citet{blo12} concluded that SN~2011fe did not have a
red giant companion.  Likewise, the light curve and spectrum of
SN~1999by match a sub-$M_{\rm Ch}$ detonation but not a $M_{\rm Ch}$
DDT \citep{blo18}.  For similar reasons, the peculiar SN ZTF18aaqeasu
seems to have been a double detonation of a sub-$M_{\rm Ch}$ WD
\citep{de19}.  Furthermore, hydrodynamical models of observed SN~Ia
light curves show that at least some progenitors must be below $M_{\rm
  Ch}$ \citep{gol18}, but with a WD mass function that peaks toward
$M_{\rm Ch}$ \citep{sca14}.  It is important to note that these
studies are not necessarily discordant because SNe~Ia could explode
through multiple channels \citep[e.g.][]{man06}.  Nonetheless, the
purpose of our study is to identify the {\it dominant} channel of
SNe~Ia in dwarf galaxies.

Nucleosynthesis predictions exist for most of these classes of models.
While the nucleosynthesis distinction between single- and
double-degenerate models may not be large enough to distinguish with
current observational data, the mass of the exploding WD has a large
effect on the production of certain elements, such as Mn and Ni
\citep[e.g.,][]{sei13b}.  Therefore, one way to address the nature of
SNe~Ia is to measure elemental abundances in SNe~Ia, their remnants,
the gas that they pollute, and the stars that form from that polluted
gas.

It is difficult to measure elemental abundances in the spectra of
SNe~Ia because the material is optically thick and possibly highly
inhomogeneous \citep{pos14}.  Although it is possible to measure the
abundances of some elements in SN~Ia remnants
\citep[e.g.,][]{bad06,bor10,bor13,lop15,gre17,mar17}, that technique
is limited to a small number of very recent explosions, and it is
limited to the few elements that can be observed with X-ray
spectroscopy.  One of the more interesting cases is the SN~Ia remnant
3C~397, which shows enhancements of Fe-peak elements
\citep{yam15,dav17}, suggesting that the progenitor was a near-$M_{\rm
  Ch}$ WD\@.

An alternative method of quantifying SN~Ia nucleosynthesis is galactic
archaeology.  SNe~Ia that exploded long ago left a chemical imprint on
the surrounding gas and stars.  We can compare the amounts of key
elements, like Mn, Fe, and Ni, in those stars to different classes of
SN~Ia models.  The Milky Way (MW) is the site of most galactic
archaeological studies of SN physics.  It is also common to compare
theoretical predictions of certain models to the abundances a single
star: the Sun \citep[e.g.,][]{mae10,sei13b}

Dwarf galaxies provide an attractive alternative to the MW for
studying SNe~Ia.  Dwarf galaxies have simple star formation histories
(SFHs), which makes their chemical evolution simple to interpret.  The
ratio of $\alpha$ elements (O, Mg, Si, and others) to Fe declines
steeply with increasing metallicity in dwarf galaxies
\citep{she01,ven04,kir11b}, which is the signature of a transition
from core collapse supernovae (CCSNe) to SNe~Ia
\citep[e.g.,][]{gil91}.  As we will show in Section~\ref{sec:trends},
most of the stars in dwarf galaxies were enriched predominantly by
SNe~Ia rather than CCSNe.  Furthermore, the dwarf galaxies have low
metallicity (${\rm [Fe/H]} \lesssim -1$).  As a result, any extra
neutrons in the nucleosynthesis can be attributed to the explosive
events and simmering rather than metallicity (see
Section~\ref{sec:theoryyields}).

We present a galactic archaeological approach to identifying the
nature of SNe~Ia in dwarf galaxies.  We use our published catalogs of
abundances of Mg, Si, Ca, and Fe \citep{kir10} as well as Cr, Co, and
Ni \citep[][see Section~\ref{sec:msmts}]{kir18}.  We fit a simple
bi-linear model to the evolution of each element, from which we
deduced the CCSN yield (Section~\ref{sec:trends}).  Then, we subtract
the CCSN yield to isolate the SN~Ia yield, which we compare to
different classes of SN~Ia models (Section~\ref{sec:constraints}).
Section~\ref{sec:discussion} discusses our conclusions in the context
of the broader literature of galactic archaeological studies of
SNe~Ia, and Section~\ref{sec:summary} summarizes the paper.


\section{Measurements of Iron-peak Abundances in Dwarf Galaxies}
\label{sec:msmts}

\citet{kir10} presented a catalog of Mg, Si, Ca, Ti, and Fe abundances
for 2961 red giants in eight dwarf spheroidal (dSph) satellites of the
MW\@.  \citet{kir18} measured Cr, Co, and Ni abundances for \ndsph\ of
these stars.  For the current study, we used a subset of these
catalogs to draw inferences about the progenitors of SNe~Ia.  Here, we
give a short recounting of the measurements, which \citet{kir10,kir18}
described in detail.

The abundances in the catalog come from spectra obtained with DEIMOS
\citep{fab03} on the Keck~II telescope.  The spectral range was
approximately 6300--9100~\AA, and the spectral resolution was
$\Delta\lambda \sim 1.2$~\AA\ (FWHM)\@.  The measurements were
obtained with spectral synthesis.  The syntheses themselves were
computed with MOOG \citep{sne73} under the approximation of local
thermodynamic equilibrium (LTE).  As described by \citet{kir18}, we
used the solar abundances of \citet{asp09} except for Mg and Fe.
Instead, we used $12 + \log [n({\rm Mg})/n({\rm H})] = 7.58$ and $12 +
\log [n({\rm Fe})/n({\rm H})] = 7.52$.

We compare these observed abundances to theoretical predictions of the
explosions of SNe~Ia in Section~\ref{sec:theoryyields}.  Therefore, it
is important to attempt to place the observed abundances on an
absolute scale as closely as possible.  To this end, \citet{kir18}
experimented with non-LTE (NLTE) corrections to the abundances.
However, they found that attempts at NLTE corrections actually
resulted in decreased abundance accuracy.  The accuracy was assessed
by computing the dispersion of Cr and Co abundances within individual
globular clusters, which are not expected to show any such dispersion.
Careful attempts to apply NLTE corrections increased the dispersion
from the LTE abundances.  This outcome likely reflects the method in
which the atmospheric parameters were determined because the NLTE
corrections usually decrease this dispersion \citep{ber10,berces10}.
Therefore, we used LTE abundances in this work.  \citet{kir18} used
the globular cluster diagnostic to quantify the systematic error in
the abundances, which includes some of the error imposed by assuming
LTE\@.  These errors are folded into the present work.

Our catalog contains only measurements with estimated uncertainties
less than 0.3~dex.  The stars in the catalog pass galaxy membership
cuts, described by \citet{kir18}, on the basis of surface gravity
(inferred from the strength of the Na~{\sc i} 8190 doublet) and radial
velocity.  In addition to these previous membership criteria, we
applied a membership criterion based on {\it Gaia} proper motions
\citep{gaiadr2}.  First, we cross-matched our catalog with {\it Gaia}.
Then, we calculated the mean proper motion in R.A.\ and
decl.\ weighted by the inverse square of the uncertainties.  Finally,
we excluded stars that were more than 3$\sigma$ discrepant from this
mean value.  This procedure excluded between one and five stars for
each dSph except Fornax, for which 22 stars were excluded.

We restricted our analysis to seven dSphs.  In order of decreasing
stellar mass, they are Fornax, Leo~I, Sculptor, Leo~II, Draco,
Sextans, and Ursa Minor.  We reserve discussion of Fornax and Leo~I
until Section~\ref{sec:discussion} because they show qualitatively
different evolution of Fe-peak elements from the other five dSphs.  As
a result, the chemical evolution model we present in
Section~\ref{sec:model} is too simple to capture their abundance
distributions.  Although Canes Venatici~I was part of the
\citet{kir10,kir18} catalogs, we do not discuss it at all because its
sample size is too small to draw meaningful conclusions.


\section{Trends of Iron-peak Abundances in Dwarf Galaxies}
\label{sec:trends}

\subsection{A Simple Model of Chemical Evolution}
\label{sec:model}

\begin{figure}
\centering
\includegraphics[width=\linewidth]{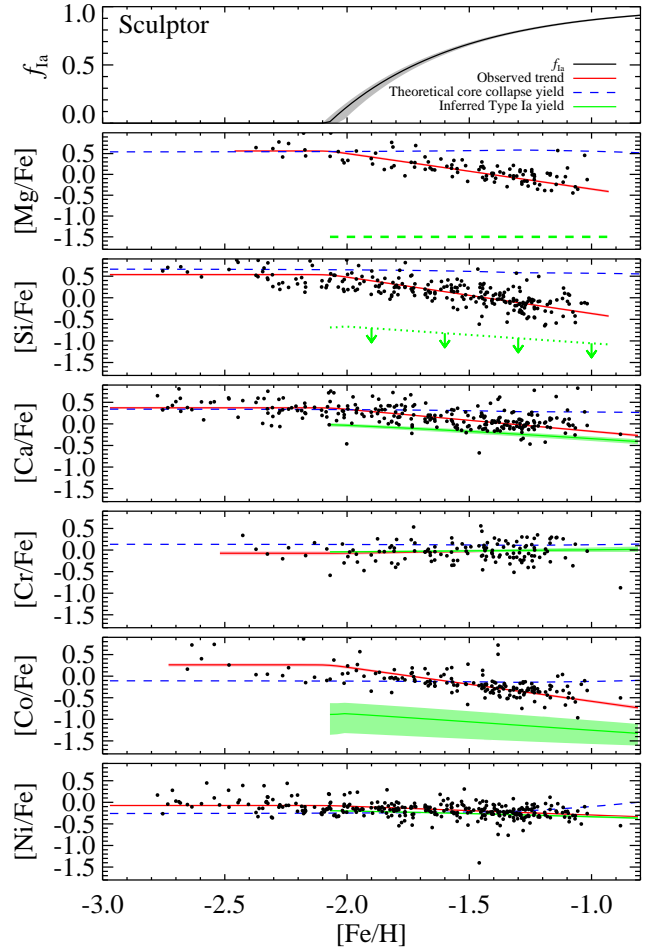}
\caption{Mg, Si, Ca, Cr, Co, and Ni abundance ratios, with respect to
  Fe, as observed in Sculptor (black points).  A simple model
  (Equation~\ref{eq:fit}, red lines) was fit to the abundances.  The
  [Mg/Fe] ratio was used to estimate the fraction of Fe ($f_{\rm Ia}$,
  top panel) that originated in SNe~Ia.  Section~\ref{sec:obsyields}
  describes a method for isolating the yield ratios (e.g., ${\rm
    [Ni/Fe]}_{\rm Ia}$) for SNe~Ia, which are shown as green curves.
  Shaded bands represent the 68\% confidence intervals.  Upper limits
  represent 95\% confidence in cases where the lower bounds of the
  yields were not well constrained.  The dashed blue curves show the
  metallicity-dependent, initial mass function-averaged yields
  predicted for CCSNe \citep{nom06}.  The dashed green line in the Mg
  panel shows the yield we assumed for SNe~Ia.\label{fig:scl}}
\end{figure}

\begin{figure}
\centering
\includegraphics[width=\linewidth]{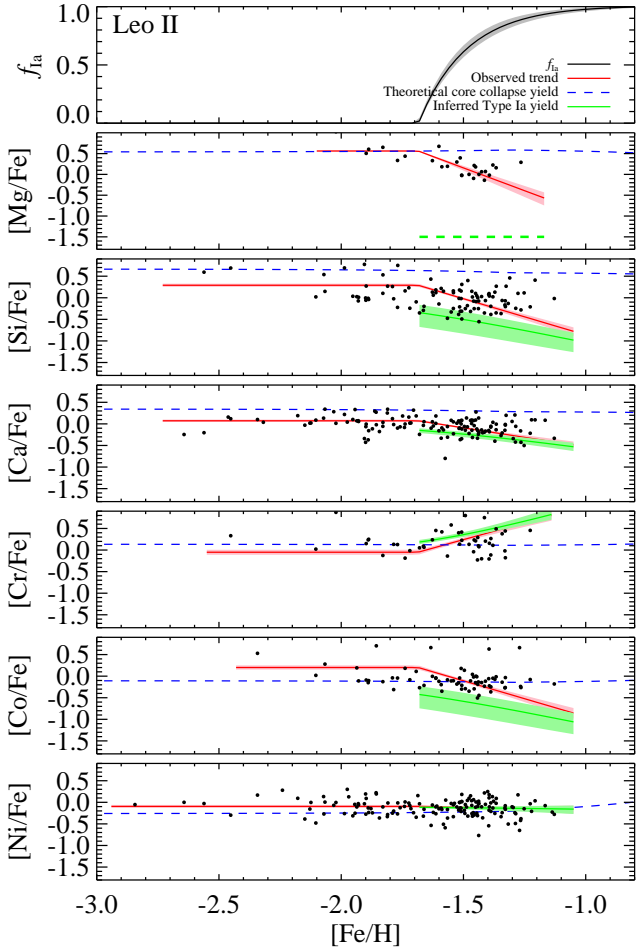}
\caption{Same as Figure~\ref{fig:scl} but for Leo~II\@.\label{fig:leoii}}
\end{figure}

\begin{figure}
\centering
\includegraphics[width=\linewidth]{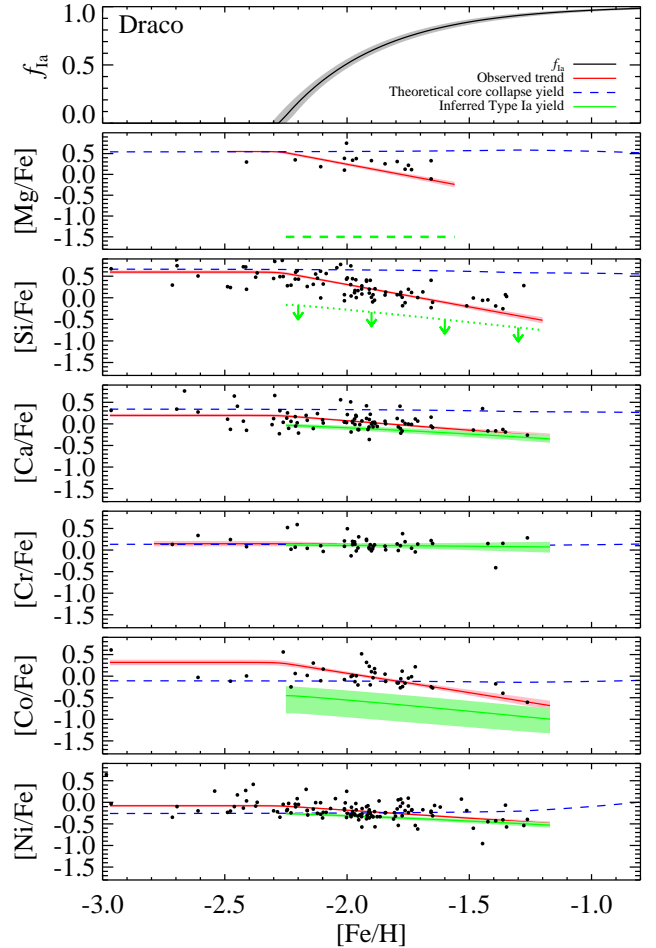}
\caption{Same as Figure~\ref{fig:scl} but for Draco\@.\label{fig:dra}}
\end{figure}

\begin{figure}
\centering
\includegraphics[width=\linewidth]{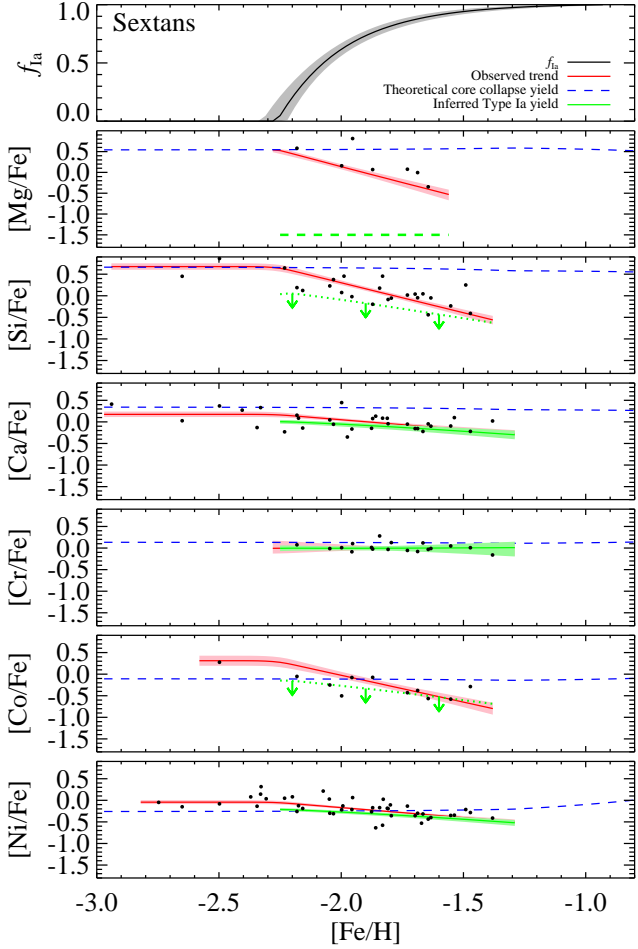}
\caption{Same as Figure~\ref{fig:scl} but for Sextans\@.\label{fig:sex}}
\end{figure}

\begin{figure}
\centering
\includegraphics[width=\linewidth]{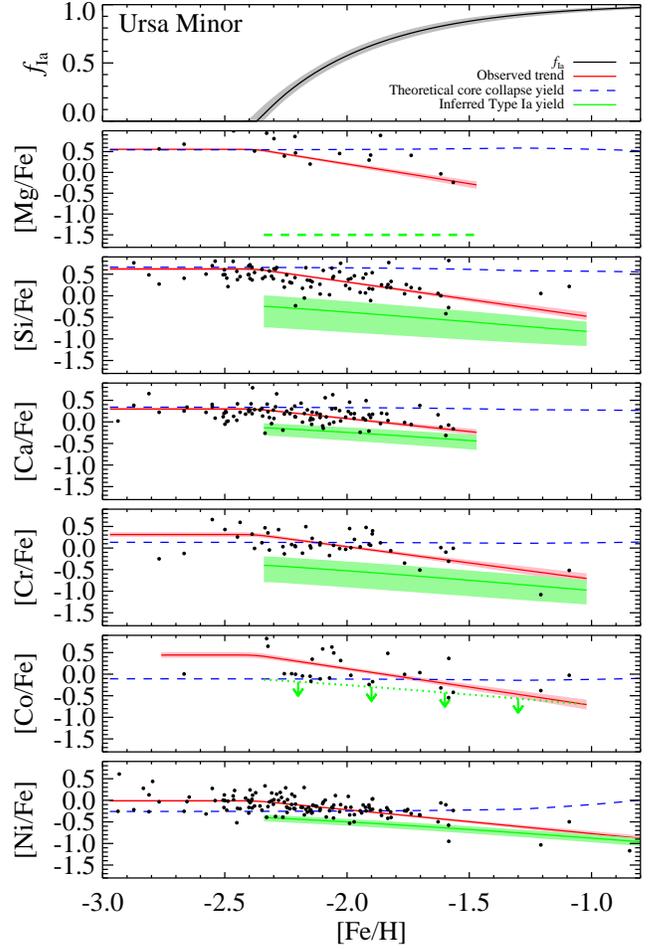}
\caption{Same as Figure~\ref{fig:scl} but for Ursa Minor\@.\label{fig:umi}}
\end{figure}

The chemical evolution of a galaxy is often expressed as
[X/Fe]\footnote{We use ``bracket notation'': ${\rm [X/Fe]} = \log
  \frac{n({\rm X})/n({\rm Fe})}{n_{\sun}({\rm X})/n_{\sun}({\rm Fe})}$
  where $n$ is atomic number density in the star's atmosphere and
  $n_{\sun}$ is the number density in the Sun's atmosphere.} versus
[Fe/H], where [X/Fe] is the ratio between the abundance of an element
X and the abundance of Fe.  The iron abundance, [Fe/H], is a proxy for
the total abundance of all metals.  The expected first-order behavior
of [X/Fe] is constant at low metallicity but sloped---either
positively or negatively---above some threshold value of [Fe/H]
\citep{whe89,gil91}.

This expectation is based on the assumption that CCSNe are the only
nucleosynthetic sources active at early times.  (However, there may be
some prompt SNe~Ia that violate this assumption, as discussed by
\citealt{man06}.)  Under the approximation (addressed in
Section~\ref{sec:ccyields}) that the CCSN yields are independent of
metallicity, [X/Fe] will be constant as [Fe/H] increases up to a
certain threshold metallicity, called ${\rm [Fe/H]}_{\rm Ia}$.  After
a delay time, corresponding to ${\rm [Fe/H]}_{\rm Ia}$, SNe~Ia begin
to explode.  They produce elements in a ratio [X/Fe] that may differ
from CCSNe.  As a result, [X/Fe] can begin to change as a function of
metallicity at ${\rm [Fe/H]} > {\rm [Fe/H]}_{\rm Ia}$.  The slope will
be negative if SNe~Ia produce a smaller ratio of [X/Fe] than CCSNe.

Figures~\ref{fig:scl} through \ref{fig:umi} show the evolution of
several abundance ratios, [X/Fe], as a function of [Fe/H] for five of
the dSphs in our sample.  The elements (X) are the $\alpha$ elements
Mg, Si, and Ca, and the Fe-peak elements Cr, Co, and Ni.  The behavior
of [X/Fe] follows the approximate pattern described above.  For
example, [Ca/Fe] in Sculptor (central panel in Figure~\ref{fig:scl})
is flat until ${\rm [Fe/H]} \approx -2.4$ and then declines nearly
linearly with increasing [Fe/H]\@.

We modeled this behavior as a constant ${\rm [X/Fe]}_{\rm CC}$ until
${\rm [Fe/H]}_{\rm Ia}$ followed by a sloped line thereafter.  The
value of ${\rm [X/Fe]}_{\rm CC}$ is the result of all of the products
of early explosions, including the CCSN yields summed over the initial
mass function (IMF) and any prompt SNe~Ia.  Instead of parameterizing
the line by a slope and intercept, we used an angle ($\theta$) and a
perpendicular offset ($b_{\perp}$).  This allowed us to use a uniform
prior on $\theta$, which avoids the preference for shallow slopes when
using a uniform prior on the slope \citep[see][]{hog10}.

\begin{equation}
   {\rm [X/Fe]}= 
\begin{cases}
    {\rm [X/Fe]}_{\rm CC}               & \text{if } {\rm [Fe/H]} \le {\rm [Fe/H]}_{\rm Ia} \\
    {\rm [Fe/H]}\,\tan \theta + \frac{b_{\perp}}{\cos \theta} & \text{if } {\rm [Fe/H]} > {\rm [Fe/H]}_{\rm Ia}
\end{cases}
\label{eq:fit}
\end{equation}

\noindent Enforcing continuity of [X/Fe] at ${\rm [Fe/H]}_{\rm Ia}$
gives the following equation for $b_{\perp}$ in terms of ${\rm
  [X/Fe]}_{\rm CC}$, ${\rm [Fe/H]}_{\rm Ia}$, and $\theta$:

\begin{equation}
b_{\perp} = {\rm [X/Fe]}_{\rm CC} \cos \theta - {\rm [Fe/H]}_{\rm Ia} \sin \theta \label{eq:bperp}
\end{equation}

We fit the chemical evolution in each dSph with Equation~\ref{eq:fit}
using maximum likelihood.  There are 13 free parameters for each dSph:
${\rm [Fe/H]}_{\rm Ia}$ and one pair of $\theta$ and $b_{\perp}$ for
each of the six [X/Fe] ratios.  Following \citet{hog10}, the
likelihood function is given by the following equations.  For a star
identified by index $i$ with ${\rm [Fe/H]}_i \le {\rm [Fe/H]}_{\rm
  Ia}$, the likelihood is relatively simple:

\begin{equation}
  L_i = \frac{1}{\sqrt{2 \pi} \delta {\rm [X/Fe]}_i} \exp \left(-\frac{\left({\rm [X/Fe]}_i - {\rm [X/Fe]}_{\rm CC}\right)^2}{2\,\delta {\rm [X/Fe]}_i^2}\right) \label{eq:Llo}
\end{equation}

\noindent where $\delta {\rm [X/Fe]}_i$ represents the measurement
uncertainty for star $i$.  For a star with ${\rm [Fe/H]}_i > {\rm
  [Fe/H]}_{\rm Ia}$, the likelihood is

\begin{eqnarray}
  L_i &=& \frac{1}{\sqrt{2 \pi} \Sigma_i} \exp \left(-\frac{\Delta_i^2}{2\Sigma_i^2}\right) \label{eq:Lhi} \\
  \Delta_i &=& {\rm [X/Fe]}_i \cos \theta - {\rm [Fe/H]}_i \sin \theta - b_{\perp} \label{eq:delta} \\
  \Sigma_i^2 &=& \delta {\rm [X/Fe]}_i^2 \cos^2 \theta + \delta {\rm [Fe/H]}_i^2 \sin^2 \theta \label{eq:sigma}
\end{eqnarray}

We imposed a prior on ${\rm [Mg/Fe]}_{\rm CC}$.  The purpose of the
prior was to ensure that the description of chemical evolution
conforms to CCSN yields for Mg.  We used predicted yields averaged
over the \citet{sal55} IMF from \citet[][indicated by a subscripted
  ``N06'']{nom06}.  The mathematical form of the prior is

\begin{equation}
  P = \frac{1}{\sqrt{2 \pi} \sigma_{\rm Mg}} \exp \left(-\frac{\left({\rm [Mg/Fe]}_{\rm N06}-{\rm [Mg/Fe]}_{\rm CC}\right)^2}{2\sigma_{\rm Mg}^2}\right) \label{eq:mgprior}
\end{equation}

\noindent The prior, $P$, is multiplied into the likelihood, $L =
\prod_i L_i$.  The value of $\sigma_{\rm Mg}$ sets the strength of the
prior.  Smaller values lead to a stronger prior, forcing better
agreement with the predicted yields.  We chose $\sigma_{\rm Mg} =
0.01$, which we consider to be a strong prior (see
Section~\ref{sec:ccyields} and Figure~\ref{fig:IIcompare}).

The fitting method was Markov chain Monte Carlo (MCMC)\@.  For
computational expediency, we maximized $\ln L$ rather than $L$.  The
chain had $3.3 \times 10^5$ links.  The first $3 \times 10^4$ links
(commonly called ``burn-in'') were discarded.  We imposed uniform
priors on all free parameters ($\theta$ and $b_{\perp}$ for each
element, as well as ${\rm [Fe/H]}_{\rm Ia}$).  We imposed an
additional prior on a value called ${\rm (X/Fe)}_{\rm Ia}$, described
in Section~\ref{sec:obsyields}.  The initial value of ${\rm
  [Fe/H]}_{\rm Ia}$ was $-2.0$, which is approximately the value at
which [$\alpha$/Fe] begins to decline in most of the dSphs in our
sample.  The initial values of $\theta$ and $b_{\perp}$ for each
element were given by a simple linear fit to the trend of [X/Fe]
versus [Fe/H] for ${\rm [Fe/H]} > -2.1$.  The proposal density for
successive links in the chain was based on the Metropolis algorithm.
The typical perturbation of ${\rm [Fe/H]}_{\rm Ia}$ and $b_{\perp}$
was 0.02~dex, and the typical perturbation of $\theta$ was
$0.2^{\circ}$.  The MCMC sampled the posterior distribution of these
parameters.  We quote the median (50$^{\rm th}$ percentile) value for
each quantity.  The asymmetric error bars are 68\% confidence
intervals around the median.

\begin{deluxetable*}{lccccc}
\tablewidth{0pt}
\tablecolumns{6}
\tablecaption{Observationally Inferred Yields\label{tab:yields_obs}}
\tablehead{\colhead{Parameter} & \colhead{Scl  } & \colhead{LeoII} & \colhead{Dra  } & \colhead{Sex  } & \colhead{UMi  }}
\startdata
${\rm [Fe/H]}_{\rm Ia}$ & $-2.12_{-0.06}^{+0.01}$ & $-1.70_{-0.01}^{+0.00}$ & $-2.36_{-0.06}^{+0.03}$ & $-2.36_{-0.03}^{+0.01}$ & $-2.42_{-0.01}^{+0.00}$ \\
$\theta({\rm Mg})$ & $-42.9_{-0.6}^{+0.4}$ & $-70.4_{-1.4}^{+0.8}$ & $-52.3_{-1.6}^{+0.7}$ & $-63.1_{-1.7}^{+1.1}$ & $-48.9_{-1.7}^{+0.9}$ \\
$b_{\perp}({\rm Mg})$ & $-0.97_{-0.02}^{+0.01}$ & $-1.40_{-0.03}^{+0.02}$ & $-1.45_{-0.05}^{+0.02}$ & $-1.75_{-0.04}^{+0.02}$ & $-1.41_{-0.06}^{+0.03}$ \\
${\rm [Mg/Fe]}_{\rm CC}$ & $+0.56 \pm 0.01$ & $+0.56 \pm 0.01$ & $+0.55 \pm 0.01$ & $+0.55 \pm 0.01$ & $+0.55 \pm 0.01$ \\
$\theta({\rm Si})$ & $-42.4_{-0.6}^{+0.4}$ & $-63.6_{-1.1}^{+0.6}$ & $-50.0_{-1.1}^{+0.6}$ & $-59.7_{-1.6}^{+0.8}$ & $-43.4_{-1.2}^{+0.7}$ \\
$b_{\perp}({\rm Si})$ & $-0.97 \pm 0.01$ & $-1.37_{-0.02}^{+0.01}$ & $-1.33_{-0.03}^{+0.02}$ & $-1.56_{-0.03}^{+0.02}$ & $-1.15_{-0.04}^{+0.02}$ \\
${\rm [Si/Fe]}_{\rm CC}$ & $+0.54 \pm 0.02$ & $+0.29 \pm 0.04$ & $+0.60 \pm 0.04$ & $+0.68_{-0.07}^{+0.08}$ & $+0.62 \pm 0.03$ \\
${\rm [Si/Fe]}_{\rm Ia}$ & $<-0.86$ & $-0.50_{-0.33}^{+0.18}$ & $<-0.57$ & $<-0.52$ & $-0.60_{-0.42}^{+0.24}$ \\
$\theta({\rm Ca})$ & $-30.1_{-0.9}^{+0.5}$ & $-48.5_{-1.7}^{+1.0}$ & $-30.2_{-1.8}^{+0.9}$ & $-35.4_{-3.3}^{+2.0}$ & $-37.5_{-2.0}^{+1.0}$ \\
$b_{\perp}({\rm Ca})$ & $-0.69_{-0.02}^{+0.01}$ & $-1.20_{-0.03}^{+0.02}$ & $-0.94_{-0.05}^{+0.03}$ & $-1.10_{-0.09}^{+0.05}$ & $-1.18_{-0.06}^{+0.03}$ \\
${\rm [Ca/Fe]}_{\rm CC}$ & $+0.37 \pm 0.02$ & $+0.07 \pm 0.03$ & $+0.20 \pm 0.04$ & $+0.17_{-0.06}^{+0.07}$ & $+0.30 \pm 0.03$ \\
${\rm [Ca/Fe]}_{\rm Ia}$ & $-0.17_{-0.05}^{+0.04}$ & $-0.24_{-0.07}^{+0.06}$ & $-0.24 \pm 0.07$ & $-0.22_{-0.09}^{+0.08}$ & $-0.43_{-0.20}^{+0.14}$ \\
$\theta({\rm Cr})$ & $ -2.9_{-1.9}^{+1.0}$ & $+48.1_{-4.2}^{+1.8}$ & $-16.9_{-4.3}^{+1.8}$ & $-29.0_{-8.1}^{+3.6}$ & $-43.0_{-1.5}^{+0.9}$ \\
$b_{\perp}({\rm Cr})$ & $-0.11_{-0.05}^{+0.03}$ & $+1.26_{-0.07}^{+0.03}$ & $-0.44_{-0.14}^{+0.06}$ & $-0.87_{-0.21}^{+0.11}$ & $-1.35_{-0.04}^{+0.02}$ \\
${\rm [Cr/Fe]}_{\rm CC}$ & $-0.07_{-0.05}^{+0.04}$ & $-0.05 \pm 0.06$ & $+0.15 \pm 0.06$ & $-0.01_{-0.13}^{+0.18}$ & $+0.31 \pm 0.05$ \\
${\rm [Cr/Fe]}_{\rm Ia}$ & $-0.02 \pm 0.03$ & $+0.36_{-0.05}^{+0.06}$ & $+0.09_{-0.09}^{+0.08}$ & $+0.00_{-0.14}^{+0.10}$ & $-0.75_{-0.37}^{+0.23}$ \\
$\theta({\rm Co})$ & $-41.2_{-0.7}^{+0.4}$ & $-64.0_{-1.1}^{+0.7}$ & $-48.0_{-1.4}^{+0.9}$ & $-59.1_{-2.1}^{+1.2}$ & $-46.1_{-1.7}^{+0.8}$ \\
$b_{\perp}({\rm Co})$ & $-1.13_{-0.02}^{+0.01}$ & $-1.41_{-0.02}^{+0.01}$ & $-1.43_{-0.04}^{+0.02}$ & $-1.72_{-0.04}^{+0.02}$ & $-1.36_{-0.05}^{+0.02}$ \\
${\rm [Co/Fe]}_{\rm CC}$ & $+0.26 \pm 0.04$ & $+0.20 \pm 0.05$ & $+0.32 \pm 0.06$ & $+0.31 \pm 0.12$ & $+0.44 \pm 0.06$ \\
${\rm [Co/Fe]}_{\rm Ia}$ & $-1.06_{-0.40}^{+0.25}$ & $-0.58_{-0.32}^{+0.20}$ & $-0.81_{-0.37}^{+0.24}$ & $<-0.60$ & $<-0.47$ \\
$\theta({\rm Ni})$ & $-13.8_{-0.7}^{+0.4}$ & $-24.1_{-4.4}^{+2.4}$ & $-25.0_{-1.8}^{+0.8}$ & $-32.4_{-2.5}^{+1.3}$ & $-32.6_{-1.1}^{+0.6}$ \\
$b_{\perp}({\rm Ni})$ & $-0.55_{-0.02}^{+0.01}$ & $-0.73_{-0.10}^{+0.06}$ & $-1.01_{-0.05}^{+0.02}$ & $-1.20_{-0.06}^{+0.03}$ & $-1.27_{-0.03}^{+0.02}$ \\
${\rm [Ni/Fe]}_{\rm CC}$ & $-0.07 \pm 0.01$ & $-0.10_{-0.03}^{+0.04}$ & $-0.08 \pm 0.02$ & $-0.04_{-0.04}^{+0.05}$ & $-0.01 \pm 0.02$ \\
${\rm [Ni/Fe]}_{\rm Ia}$ & $-0.26_{-0.02}^{+0.01}$ & $-0.13_{-0.04}^{+0.03}$ & $-0.43_{-0.05}^{+0.04}$ & $-0.44_{-0.06}^{+0.05}$ & $-0.68_{-0.09}^{+0.07}$ \\
\enddata
\tablecomments{All values of $\theta$ are given in degrees.  All other values are dex relative to the Sun.  The values of ${\rm [X/Fe]}_{\rm Ia}$ are evaluated at ${\rm [Fe/H]} = -1.5$.  Errors represent the 68\% confidence intervals.  Asymmetric errors are quoted where the upper and lower errors differ.  In cases where ${\rm (X/Fe)}_{\rm Ia}$ was consistent with zero, upper limits on ${\rm [X/Fe]}_{\rm Ia}$ represent 95\% confidence.}
\end{deluxetable*}

The best-fit models are shown as red lines in Figures~\ref{fig:scl}
through \ref{fig:umi}.  The pink regions show the 68\% confidence
intervals.  Table~\ref{tab:yields_obs} provides the 13 best-fit
parameters for each dSph.  It also provides the derived quantities
${\rm [X/Fe]}_{\rm Ia}$ and ${\rm [X/Fe]}_{\rm CC}$, which are
described in Sections~\ref{sec:obsyields} and \ref{sec:ccyields}.

\subsection{Observationally Inferred Yields of SNe Ia}
\label{sec:obsyields}

The amount of an element in a star is a combination of the various
nucleosynthetic sources that contributed to its birth cloud.  These
sources include SNe, winds from intermediate- and low-mass stars, and
even neutron star mergers.  The major contributors to the elements
considered in this work (Mg, Si, Ca, Cr, Fe, Co, and Ni) are CCSNe and
SNe~Ia.

The amount of an element X in a star can be represented as a sum of
the contribution from both types of SN:

\begin{equation}
{\rm X}_* = {\rm X}_{\rm CC} + {\rm X}_{\rm Ia} \label{eq:x}
\end{equation}

\noindent We have implicitly assumed that element X has a negligible
contribution from sources other than SNe and that the abundance
of element X has not changed since the star's birth.  We can represent
the ratio of two elements in the star as a ratio of their
contributions from both types of SNe.  Iron is typically used
as the comparison element.  Bracket notation is not used in the
following equations because the element ratios, e.g., (X/Fe), are
linear, not logarithmic:

\begin{equation}
\left( \frac{{\rm X}}{{\rm Fe}} \right)_* = \frac{{\rm X}_{\rm CC} + {\rm X}_{\rm Ia}}{{\rm Fe}_{\rm CC} + {\rm Fe}_{\rm Ia}} \label{eq:xfe}
\end{equation}

We now define a ratio, $R$, of the amount of iron that comes from
SNe~Ia compared to the amount that comes from CCSNe.

\begin{equation}
R \equiv \frac{{\rm Fe}_{\rm Ia}}{{\rm Fe}_{\rm CC}} \label{eq:rdef}
\end{equation}

\noindent Equation~\ref{eq:xfe} can be expressed in terms of $R$ by
dividing the numerator and denominator on the right side by ${\rm
  Fe}_{\rm CC}$:

\begin{eqnarray}
\left( \frac{{\rm X}}{{\rm Fe}} \right)_* &=& \frac{({\rm X/Fe})_{\rm CC} + {\rm X}_{\rm Ia}/{\rm Fe}_{\rm CC}}{1 + R} \\
 &=& \frac{({\rm X/Fe})_{\rm CC} + R({\rm X/Fe})_{\rm Ia}}{1 + R} \label{eq:xfer}
\end{eqnarray}

\noindent Equation~\ref{eq:xfer} can then be solved for $R$:

\begin{equation}
R = \frac{({\rm X/Fe})_{\rm CC} - ({\rm X/Fe})_*}{({\rm X/Fe})_* - ({\rm X/Fe})_{\rm Ia}} \label{eq:r}
\end{equation}
  
\noindent Alternatively, Equation~\ref{eq:xfer} can be solved for
$({\rm X/Fe})_{\rm Ia}$:

\begin{equation}
\left( \frac{{\rm X}}{{\rm Fe}} \right)_{\rm Ia} = \frac{R+1}{R}\left( \frac{{\rm X}}{{\rm Fe}} \right)_* - \frac{1}{R}\left( \frac{{\rm X}}{{\rm Fe}} \right)_{\rm CC} \label{eq:xfeia}
\end{equation}

As an aside, we can also write expressions for the fractions of iron
that came from SNe~Ia ($f_{\rm Ia}$) or CCSNe ($f_{\rm CC}$):

\begin{eqnarray}
f_{\rm Ia} &\equiv& \frac{{\rm Fe}_{\rm Ia}}{{\rm Fe}_{\rm CC} + {\rm Fe}_{\rm Ia}} = \frac{R}{R+1} \nonumber \\
         &=& \frac{({\rm X/Fe})_{\rm CC} - ({\rm X/Fe})_*}{({\rm X/Fe})_{\rm CC} - ({\rm X/Fe})_{\rm Ia}} \label{eq:fIa} \\
f_{\rm CC} &\equiv& \frac{{\rm Fe}_{\rm CC}}{{\rm Fe}_{\rm CC} + {\rm Fe}_{\rm Ia}} = \frac{1}{R+1} \nonumber \\
         &=& \frac{({\rm X/Fe})_* - ({\rm X/Fe})_{\rm Ia}}{({\rm X/Fe})_{\rm CC} - ({\rm X/Fe})_{\rm Ia}} \label{eq:fCC}
\end{eqnarray}

The ratio $R$ is defined only in terms of iron.  Thus,
Equation~\ref{eq:r} can be used to determine $R$ using one element,
${\rm X}_1$.  The resulting value of $R$ is general for any other
element, ${\rm X}_2$, as long as both ${\rm X}_1$ and ${\rm X}_2$ can
be considered to come exclusively from SNe.  Once $R$ is known,
Equation~\ref{eq:xfeia} can be used to infer the SN~Ia yield for the
ratio $({\rm X}/{\rm Fe})_{\rm Ia}$ for any element X\@.

The ideal element ${\rm X}_1$ for solving Equation~\ref{eq:r} is one
that has well-known yields from both CCSNe and SNe~Ia and is also
measured in our spectroscopic sample.  Magnesium is the element that
best satisfies these criteria; it is synthesized almost exclusively in
CCSNe.  Nearly all models of SNe~Ia agree that virtually no Mg is
produced.

The ratio $R$ can be calculated from $({\rm X/Fe})_{\rm CC}$, $({\rm
  X/Fe})_{\rm Ia}$, and $({\rm X/Fe})_*$.  We treated the first two
quantities as constants, but $({\rm X/Fe})_*$ varies from star to
star.  In other words, $R$ is a function of time.  As in
Section~\ref{sec:trends}, we used [Fe/H] as a proxy for time because
time is not directly observable.

In principle, $R$ could be calculated for individual stars.  However,
this measurement would be noisy.  In some cases, it could even lead to
negative (unphysical) values for $R$.  Therefore, we used the
bi-linear chemical evolution model (Equation~\ref{eq:fit}) to average
over the noise of individual measurements.

The model fits directly gave ${\rm [X/Fe]}_{\rm CC}$ and ${\rm
  [X/Fe]}_*$ but not ${\rm [X/Fe]}_{\rm Ia}$.  In the case of Mg, we
adopted a value of ${\rm [Mg/Fe]}_{\rm Ia} = -1.5$.  For reference,
Table~\ref{tab:yields_theory} provides [Mg/Fe] predictions from a
variety of SN~Ia models.  The precise value is not particularly
important because the ${\rm (Mg/Fe)}_{\rm Ia}$ ratio in linear units
is very close to zero.  We have confirmed that changing the value of
${\rm [Mg/Fe]}_{\rm Ia}$ to $-1.0$ or $-2.5$ does not qualitatively
alter our inferences of ${\rm [X/Fe]}_{\rm Ia}$.  We even tried ${\rm
  [Mg/Fe]}_{\rm Ia} = -0.3$, which is the highest predicted value for
the most extreme SN~Ia model in Table~\ref{tab:yields_theory}.  In
this case, some of our lowest inferences of ${\rm [X/Fe]}_{\rm Ia}$,
including ${\rm [Si/Fe]}_{\rm Ia}$, ${\rm [Ca/Fe]}_{\rm Ia}$, and
${\rm [Co/Fe]}_{\rm Ia}$, increased noticeably.  Importantly, ${\rm
  [Ni/Fe]}_{\rm Ia}$, which is the ratio that most directly affects
our conclusions about SNe~Ia (Section~\ref{sec:Iacompare}), barely
changed.  Therefore, our results do not depend much on the exact value
of ${\rm [Mg/Fe]}_{\rm Ia}$.

With ${\rm [Mg/Fe]}_{\rm Ia}$ in hand, we had all of the variables
required to solve for $R$ as a function of [Fe/H]\@.  The top panels
of Figures~\ref{fig:scl} to \ref{fig:umi} show $f_{\rm Ia} = R/(R+1)$
for each of the dSphs in our sample.  The values of $f_{\rm Ia}$ were
calculated at each step in the MCMC chain that evaluated the
parameters of the chemical evolution model.  The widths of the $f_{\rm
  Ia}$ bands in the figures enclose 68\% of the successful MCMC
trials.

Once $f_{\rm Ia}$ (or alternatively, $R$) was known as a function of
[Fe/H] for each dSph, Equation~\ref{eq:xfeia} gave the empirical SN~Ia
yield for an arbitrary element X\@.  We already assumed a value of
${\rm [Mg/Fe]}_{\rm Ia}$ to calculate $f_{\rm Ia}$.  Therefore, it
would not have made sense to use Equation~\ref{eq:xfeia} for Mg.  On
the other hand, we could infer the SN~Ia yields of [Si/Fe], [Ca/Fe],
[Cr/Fe], [Co/Fe], and [Ni/Fe].  We did so by evaluating
Equation~\ref{eq:xfeia} during the MCMC fitting that determined the
parameters of the chemical evolution model.

Negative values of ${\rm (X/Fe)}_{\rm Ia}$ are not physical.
Therefore, we imposed a prior in the computation of the likelihood
function to enforce non-negativity.  If an iteration in the MCMC chain
yielded one or more negative values of ${\rm (X/Fe)}_{\rm Ia}$, then
the likelihood $L$ was made to be zero.

Figures~\ref{fig:scl} to \ref{fig:umi} show in green bands the
inferred yield ratios as a function of [Fe/H] for each of the dSphs in
our sample.  The widths of the bands reflect the 68\% confidence
intervals.  Table~\ref{tab:yields_obs} gives ${\rm [X/Fe]}_{\rm Ia}$
for each dSph at a reference metallicity of ${\rm [Fe/H]} = -1.5$.

In some cases, the posterior distributions of ${\rm (X/Fe)}_{\rm Ia}$
are peaked toward zero, i.e., ${\rm [X/Fe]}_{\rm Ia}$ approaches
$-\infty$.  These cases can be interpreted as upper limits.  All upper
limits are quoted at the 95\% confidence level, i.e., the value below
which 95\% of the MCMC iterations are found.

\begin{figure*}
\centering
\includegraphics[width=\linewidth]{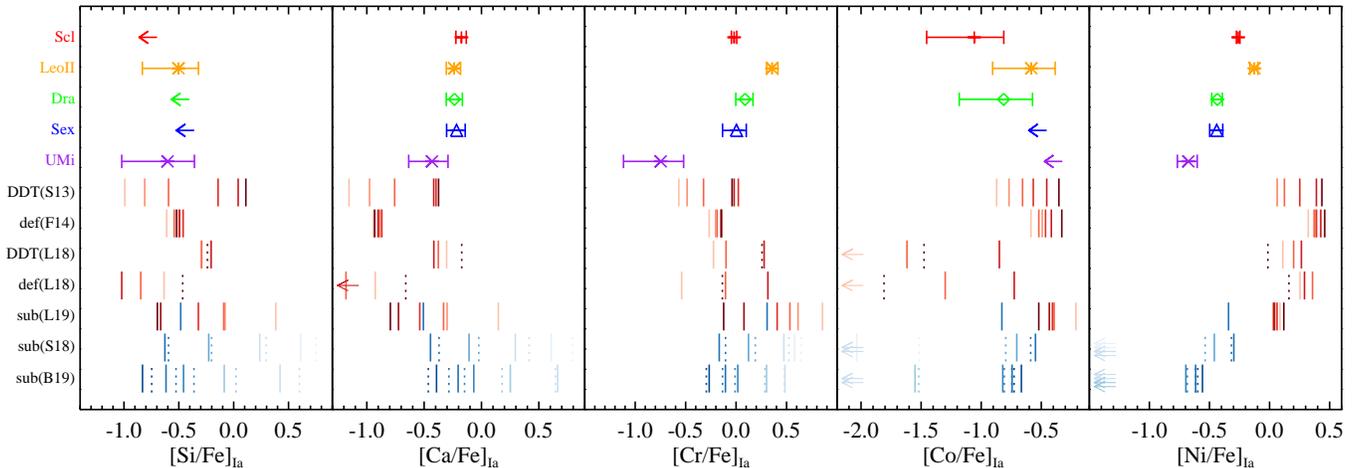}
\caption{SN~Ia yields at ${\rm [Fe/H]} = -1.5$ inferred from
  observations of dSph stars (rainbow colors) compared to
  theoretically predicted yields for various models (described in
  Table~\ref{tab:models}).  The dSphs are ordered from most massive
  (Sculptor) to least massive (Ursa Minor).  For the theoretical
  yields, shades of red indicate yields at solar metallicity, whereas
  shades of blue indicate a metallicity of $10^{-1.5}~Z_{\sun}$.
  Lighter shading indicates fewer ignition sites (S13 and F14), lower
  initial density (L18), or lower-mass WDs (L19, S18, and B19).
  Dotted lines represent special cases: WDD2 or W7 (L18), ${\rm C/O} =
  30/70$ (S18), or $\xi_{\rm CO} = 0.0$ (B19). Leftward-pointing
  arrows for the theoretical yields indicate yield ratios below the
  left limit of the plot.\label{fig:Iacompare}}
\end{figure*}

Figure~\ref{fig:Iacompare} shows the inferred SN~Ia yield ratios at
${\rm [Fe/H]} = -1.5$ for the five dSphs.  The inferences for ${\rm
  [Si/Fe]}_{\rm Ia}$, ${\rm [Ca/Fe]}_{\rm Ia}$, and ${\rm
  [Co/Fe]}_{\rm Ia}$ are consistent within $2\sigma$ for all of the
dSphs.  The formal error bars on ${\rm [Cr/Fe]}_{\rm Ia}$ and ${\rm
  [Ni/Fe]}_{\rm Ia}$ are very small for four of the dSphs, which makes
their differences appear especially significant.  However, we show in
Section~\ref{sec:constraints} that the differences do not complicate
our conclusion about the progenitor masses of SNe~Ia.  Ursa Minor's
error bars on ${\rm [X/Fe]}_{\rm Ia}$ are larger than those of the
other dSphs.  The size of the error bar results from subtracting a
large number from a large number to produce a small number.  The
fractional error bar (as presented in logarithmic space) is large for
this operation.  Interestingly, ${\rm [Ni/Fe]}_{\rm Ia}$ seems to have
a significant range among the five dSphs.

We constrained ${\rm [Si/Fe]}_{\rm Ia}$ to be a small value.  The most
constraining upper limit, provided by Sculptor, is ${\rm [Si/Fe]}_{\rm
  Ia} < \sifeiaulimscl$.  It may seem odd that we conclude that SNe~Ia
make little Si because they are classified by the presence of Si
absorption in their spectra.  However, not much Si production is
required to induce deep Si absorption features in SNe~Ia spectra.  For
example, \citet{hac13} modeled the light curves of the WDD3 and W7
explosion models of \citet{iwa99}.  Both models predict Si and Fe
production at approximately one third the solar ratio (${\rm [Si/Fe]}
= -0.5$).  This is more than sufficient to produce theoretical spectra
that closely resemble observed spectra of SNe~Ia.

\subsection{Comparison to Theoretical Yields of CCSNe}
\label{sec:ccyields}

\begin{figure*}
\centering
\includegraphics[width=\linewidth]{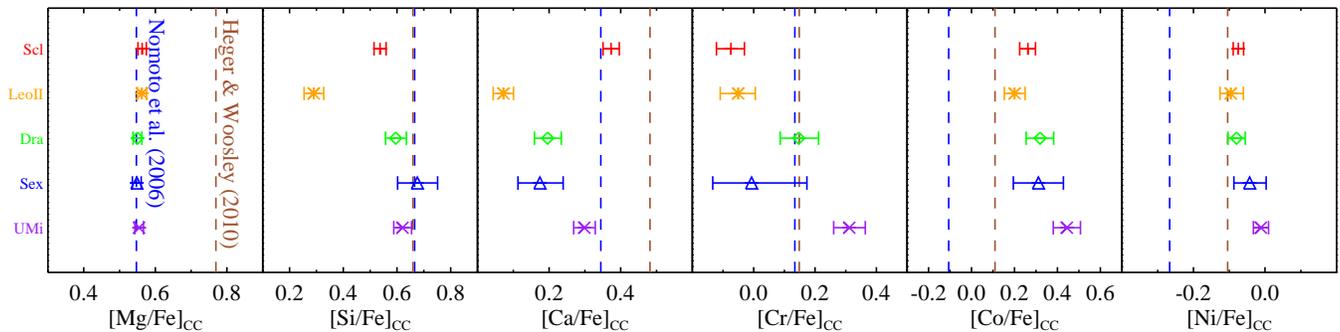}
\caption{CCSN yields inferred from observations of dSph stars compared
  to theoretically predicted yields (\citealt{nom06}, dashed blue
  line, and \citealt{heg10}, dashed brown line).\label{fig:IIcompare}}
\end{figure*}

It is instructive to compare the observationally inferred CCSN yields
with theoretical predictions from simulations of CCSNe.
Figures~\ref{fig:scl}--\ref{fig:umi} show the metallicity-dependent
predicted CCSN yields of \citet{nom06} as blue dashed lines.  The very
slight dependence on metallicity over the metallicity range of our
sample justifies our approximation that the CCSN yields are
independent of metallicity.  Figure~\ref{fig:IIcompare} compares the
CCSN yields inferred from dSphs with the predictions of \citet{nom06}
and \citet{heg10}.  Both studies used one-dimensional hydrodynamical
codes, but the explosion initiation was different.  \citet{nom06}
injected thermal energy (a ``thermal bomb''), whereas \citet{heg10}
used a piston (momentum injection).  In the case of
\citeauthor{nom06}, we used the zero-metallicity SN (not hypernova)
yields averaged over the Salpeter IMF\@.  \citet{heg10} presented only
zero-metallicity yields, but they varied many parameters in the
simulations.  We used their ``standard'' explosion model, in which a
piston at the base of the oxygen shell initiated the explosion.
Mixing followed the ``standard'' prescription, and the explosion
energy was $1.2 \times 10^{51}$~erg.  We averaged the yields over a
Salpeter IMF in the range 10--100~$M_{\sun}$.

The inferred yields generally agree with the predictions.  The left
panel of Figure~\ref{fig:IIcompare} shows that the ${\rm [Mg/Fe]}_{\rm
  CC}$ inferences fall exactly on the predictions of \citet{nom06},
but we effectively forced this agreement by imposing a strong prior on
${\rm [Mg/Fe]}_{\rm CC}$ (Equation~\ref{eq:mgprior}).  The discrepancy
between the two theoretical predictions of ${\rm [Mg/Fe]}_{\rm CC}$
raises the question whether our choice of prior \citep{nom06} affects
our conclusions.  To address this question, we re-evaluated all of the
results of this study using a different prior.  First, we note that
the observed [Mg/Fe] ratios in the stars reach as high as ${\rm
  [Mg/Fe]}_{\rm N06}$.  Therefore, it would not make sense to consider
a lower value of ${\rm [Mg/Fe]}_{\rm CC}$.  Instead, we considered
${\rm [Mg/Fe]}_{\rm CC} = +0.8$, roughly in line with \citeposs{heg10}
prediction.  While the value of ${\rm [Fe/H]}_{\rm Ia}$ did tend to
decrease when assuming a larger prior on ${\rm [Mg/Fe]}_{\rm CC}$, the
conclusions we draw in Section~\ref{sec:constraints} did not change.
In particular, the results shown in Figure~\ref{fig:Iacompare} shifted
by the size of the error bars or less.  Therefore, our conclusions on
the progenitors of SNe~Ia are mostly insensitive to reasonable choices
of the ${\rm [Mg/Fe]}_{\rm CC}$ prior.

For most galaxies, the inferences for ${\rm [Si/Fe]}_{\rm CC}$ agree
well with both sets of yields, whereas the inferences for ${\rm
  [Ca/Fe]}_{\rm CC}$ scatter around \citeauthor{nom06}'s predictions,
and the inferences for ${\rm [Cr/Fe]}_{\rm CC}$ scatter on either side
of both sets of predictions.  Leo~II is an outlier in ${\rm
  [Si/Fe]}_{\rm CC}$ and ${\rm [Ca/Fe]}_{\rm CC}$.  It is possible
that our bi-linear model is not as well suited to Leo~II's more
extended SFH compared to the other dSphs.  The inferences for ${\rm
  [Co/Fe]}_{\rm CC}$ exceed the predictions of both
\citeauthor{nom06}\ and \citeauthor{heg10}\ by several tenths of a
dex.  This discrepancy could indicate that CCSNe are more neutron-rich
than the simulations assumed, or it could indicate that we have not
properly accounted for NLTE corrections for the Co abundance
measurements \citep[see][]{kir18}.  Interestingly, the inferences for
${\rm [Ni/Fe]}_{\rm CC}$ agree well with those of
\citeauthor{heg10}\ but less well with those of \citeauthor{nom06}

\subsection{Discussion of the Model}
\label{sec:modeldiscussion}

The ``kinked line'' description of chemical evolution in
Section~\ref{sec:model} is simplistic.  First, the chemical evolution
depends sensitively on the star formation rate (SFR)\@.  Any increase
in SFR, such as a revived burst of star formation, will be accompanied
by an increase in the ratio of CCSNe to SNe~Ia, which will bring
[X/Fe] closer to the CCSN value.  Therefore, the details of the
shape---including deviations from a straight line---of [X/Fe] versus
[Fe/H] can be predicted only if the SFH is known.  Second, the model
assumes that SNe~Ia have an appreciable delay time.  However,
empirically derived delay time distributions indicate that the minimum
delay is much less than 420~Myr and possibly as small as 40~Myr
\citep{mao12,mao17}.  As a result, the period of chemical evolution
corresponding to CCSN-only nucleosynthesis may be so short as to be
irrelevant.  In this case, any changes in the slope of [X/Fe] with
[Fe/H] do not indicate the onset of SNe~Ia but rather a change in the
SFR\@.  Third, both CCSN and SN~Ia yields for some elements depend on
the progenitor's metallicity \citep{woo95,nom06,pie17}.  As a result,
[X/Fe] may not be constant at low [Fe/H], even if there is a period of
chemical evolution from CCSNe only.  (However, the yields predicted by
\citealt{nom06}\ have a negligible dependence on metallicity over the
metallicity range of our sample, as discussed in
Section~\ref{sec:ccyields}.)  Fourth, the yield of a stellar
population depends on the stellar IMF\@.  Consequently, [X/Fe] depends
not only on SFR but also on the IMF\@.  Finally, the SN~Ia yields
could change with time.  For example, if SNe~Ia explode at masses
below $M_{\rm Ch}$, the typical mass of exploding WDs could decrease
over time as WDs of lower mass begin to appear \citep[e.g.,][]{she17}.

Despite its simplicity, the ``kinked line'' model of chemical
evolution still provides a good estimate of the ratio of CCSNe to
SNe~Ia, provided that the galaxy's gas is well mixed at all times.
\citet{kir11b} and \citet{esc18} showed that this assumption is
reasonable for ancient dwarf galaxies (smaller than Fornax) by
demonstrating that [$\alpha$/Fe] has little dispersion at a given
[Fe/H]\@.  As long as [Fe/H] can be used as a proxy for time, then the
small dispersion in [$\alpha$/Fe] indicates that the mixing timescale
is shorter than the star formation timescale.  As a result, none of
the shortcomings of the ``kinked line'' model enumerated in the
previous paragraph leads to any ambiguity in interpreting [X/Fe] as a
linear combination of CCSNe and SNe~Ia.  We do need to assume that the
elements considered here originate exclusively in CCSNe and SNe~Ia,
which we believe to be a good assumption.

\subsection{Comparison to Prior Work}
\label{sec:priorwork}

In nearly all of the cases studied here, the elemental ratios decline
with increasing metallicity.  This behavior indicates that Fe-peak
elements are produced more slowly relative to Fe as the dwarf galaxies
evolve.

Previous studies have noted the same pattern in other dwarf galaxies.
For example, \citet{coh10} demonstrated with high-resolution spectra
that [Co/Fe] and [Ni/Fe] decline with increasing [Fe/H] in Ursa Minor.
On the other hand, they found the two lowest-metallicity stars in
their sample had [Cr/Fe] ratios significantly lower than the other
stars.  Although these two stars are in our sample, the
signal-to-noise ratio of the DEIMOS spectra is not sufficient to
measure any Fe-peak abundance other than Ni.  The trend of [Cr/Fe] for
the stars with $-2.5 < {\rm [Fe/H]} < -1.5$ is approximately flat in
both samples.

The decline of [Co/Fe] and [Ni/Fe] with increasing [Fe/H] has also
been noted in more massive galaxies, such as Fornax
\citep{let10,lem14} and Sagittarius \citep{sbo07,has17}.  After NLTE
corrections, \citet{ber10} and \citet{berces10} also measured
near-solar ratios of [Cr/Fe] and declining [Co/Fe] with increasing
metallicity in the metal-poor Galactic halo.  Although the Large
Magellanic Cloud does not seem to display these decreasing
iron-peak-to-iron ratios, the ratios are sub-solar \citep{pom08}.
This behavior is not limited to Co and Ni.  Even Zn ($Z=30$, not
measurable in our spectra) shows a downward trend of [Zn/Fe] with
increasing [Fe/H] in Sculptor and other Local Group dwarf galaxies
\citep{sku17}.

Taken as a whole, the [Co/Fe] and [Ni/Fe] ratios of the MW's dwarf
galaxies seem to follow the same pattern as [$\alpha$/Fe].  The ratios
are lower than in the MW at the same metallicity, and they decline
with metallicity in nearly all cases.  The interpretation in the
context of the above description of chemical evolution is that dwarf
galaxies experienced a larger ratio of SNeIa:CCSNe than the components
of the MW at comparable metallicity (the MW halo for smaller dwarf
galaxies and the MW disks for the Magellanic Clouds).  Furthermore,
the average SN~Ia in dwarf galaxies produces stable Co and Ni in a
lower proportion relative to Fe than the average CCSN\@.


\section{Constraints on SNe Ia}
\label{sec:constraints}

The [$\alpha$/Fe] ratios in dwarf galaxies indicate that their
chemical evolution is dominated by SNe~Ia at late times.  This
conclusion can be drawn from the low values of [O/Fe] and [Mg/Fe],
among other ratios, found in the stars with higher metallicities in
the dwarf galaxies \citep{she01,she03,ven04,kir11b}.  The dominance of
SNe~Ia makes dwarf galaxies among the best places to study the effect
of SNe~Ia on chemical evolution.  However, many dSphs, including the
five main galaxies in this paper, have short SFHs.  As a result, an
individual dSph with a specific SFH may not sample all varieties of
SNe~Ia, some of which could have delay times exceeding several Gyr.

In Section~\ref{sec:obsyields}, we used the [Mg/Fe] ratio to estimate
the amount of Fe that comes from CCSNe versus SNe~Ia in each star.  We
subtracted the CCSN contribution from the observed abundance ratio of
each Fe-peak element to estimate the ratio from SNe~Ia only.  We now
compare these observationally inferred SN~Ia yields to theoretical
predictions of SN~Ia yields.

\subsection{Theoretical Yields of SNe Ia}
\label{sec:theoryyields}

\defcitealias{sei13a}{S13}
\defcitealias{fin14}{F14}
\defcitealias{leu18}{L18}
\defcitealias{leu19}{L19}
\defcitealias{she18}{S18}
\defcitealias{bra19}{B19}

\begin{deluxetable*}{lll}
\tablewidth{0pt}
\tablecolumns{3}
\tablecaption{Type Ia Supernova Models\label{tab:models}}
\tablehead{\colhead{Model} & \colhead{Authors} & \colhead{Description}}
\startdata
DDT(\citetalias{sei13a}) & \citet{sei13a} & $M_{\rm Ch}$, 3D, DDT, multiple ignition sites \\
def(\citetalias{fin14}) & \citet{fin14} & $M_{\rm Ch}$, 3D, pure deflagration, multiple ignition sites \\
DDT(\citetalias{leu18}) & \citet{leu18} & $M_{\rm Ch}$, 2D, DDT, varying initial central density \\
def(\citetalias{leu18}) & \citet{leu18} & $M_{\rm Ch}$, 2D, pure deflagration, varying initial central density \\
sub(\citetalias{leu19}) & \citet{leu19} & sub-$M_{\rm Ch}$, 2D, double detonation with He shell \\
sub(\citetalias{she18}) & \citet{she18} & sub-$M_{\rm Ch}$, 1D, detonation of bare CO WD, two choices of C/O mass ratio \\
sub(\citetalias{bra19}) & \citet{bra19} & sub-$M_{\rm Ch}$, 1D, detonation of bare CO WD, two choices of ${}^{12}{\rm C} + {}^{16}{\rm O}$ reaction rate \\
\enddata
\end{deluxetable*}

Several nucleosynthetic predictions have been published in the last
several years for different types of SN~Ia explosions.  We considered
six recent studies that sampled a variety of explosion types.

Table~\ref{tab:models} summarizes the salient features of the
explosion simulations, which we describe in more detail below.  The
table assigns abbreviations to the studies: deflagration-to-detonation
transition (``DDT''), pure deflagration (``def''), and
sub-Chandrasekhar detonations (``sub'').  The code in parentheses
refers to the first author and year of publication.
\citet[][\citetalias{leu18}]{leu18} simulated both DDTs and pure deflagrations.  Each
of the models varied some parameters that affected the
nucleosynthesis.  We considered a subset of those models, focusing on
the models that their authors considered ``standard'' or ``benchmark''
models.  Table~\ref{tab:yields_theory} gives the [Si/Fe], [Ca/Fe],
[Cr/Fe], [Co/Fe], and [Ni/Fe] yield predictions from each those
models.  Figure~\ref{fig:Iacompare} also represents the theoretical
yields as vertical lines.

\vspace{4mm}

{\bf DDT:} We chose two sets of near-$M_{\rm Ch}$ DDT models.
Multi-dimensional simulations are particularly important for
deflagrations, where the burning front is highly textured.  Therefore,
we considered only multi-dimensional deflagration simulations.
(However, \citealt{kat19} cautioned that all multi-dimensional SN~Ia
simulations to date probably have not resolved the initiation of
detonation.)

\citet[][\citetalias{sei13a}]{sei13a} computed 3D models with multiple
off-center ignition sites in CO WDs with a central density of $\rho_c
= 2.9 \times 10^9$~g~cm$^{-3}$.  The model names specify the number of
ignition sites.  For example, N100 \citep[the original model
  of][]{rop12} has 100 ignition sites.  Most models assume that the WD
is 47.5\% ${}^{12}$C, 50\% ${}^{16}$O, and 2.5\% ${}^{22}$Ne by mass.
The amount of ${}^{22}$Ne corresponds to the amount expected in a WD
after the evolution of a solar-metallicity star.  \citetalias{sei13a}
also considered lower metallicity versions of the N100 model, which we
used in our work.  They also constructed models of low density, high
density, and a compact configuration of ignition sites, but we did not
consider them.

We also considered the DDT models of \citetalias{leu18}.  In contrast to the \citetalias{sei13a}
models, the \citetalias{leu18} models were computed in 2D, and the detonation began
at a single point in the center of the WD\@.  \citetalias{leu18} simulated a variety
of central densities.  We considered the models at $\rho_c = \{1,3,5\}
\times 10^9$~g~cm$^{-3}$.  They used the same initial composition as
\citetalias{sei13a}.  They also treated metallicity in the way described above, but
they considered a range of metallicity for most of their models.
Finally, \citetalias{leu18} updated ``WDD2,'' the classic DDT model of \citet{iwa99},
with modern electron capture rates.  We also considered that updated
model.

\vspace{4mm}

{\bf Pure deflagrations:} Our comparison set also includes two sets of
pure deflagrations of near-$M_{\rm Ch}$ WDs.  These models are often
considered to represent Type~Iax SNe \citep[e.g.,][]{kro15}.  As a
result, the simulated properties, like the light curve, spectrum, and
nucleosynthesis yields, may not be applicable to ``normal'' SNe~Ia.

\citet[][\citetalias{fin14}]{fin14} based their 3D simulations of pure deflagrations
closely on the DDT models of \citetalias{sei13a}.  The \citetalias{fin14} models also presumed
between 1 and 1600 off-center sites of ignition, but they did not
transition to detonations.  They considered only solar metallicity.

\citetalias{leu18} also computed pure deflagrations in addition to their DDT models.
The deflagration and DDT models were exploded with the same initial
values of central density and metallicity.  As they did with the WDD2
model, \citetalias{leu18} also updated the pure-deflagration ``W7'' model of
\citet{iwa99}.  We included that model in our comparison set.  W7 is
the only deflagration model considered here that is computed at a
variety of non-solar metallicities.

\vspace{4mm}

{\bf Sub-$M_{\rm Ch}$:} We also compared our inferred yields against
three sets of sub-$M_{\rm Ch}$ detonations.  \citet[][\citetalias{leu19}]{leu19} used
the same 2D code as their earlier work.  They exploded the simulated
WDs using double detonation.  The WDs were equal parts C and O by
mass.  The first detonation started in a He shell on the surface.
They simulated various He shell masses ($M_{\rm He}$).  The He
detonation shocked the interior C and O, which caused a second
detonation.  They considered WD masses in the range $0.90$ to
$1.20~M_{\sun}$.  All models were computed at solar metallicity except
the $1.10~M_{\sun}$ (``benchmark'') model, which was also computed at
metallicities ranging from $0.1$ to $5Z_{\sun}$. Metallicity was
approximated by the amount of ${}^{22}$Ne present.

\citet[][\citetalias{she18}]{she18} simulated 1D, spherically symmetric detonations
of bare CO WDs.  The detonations began at the centers of the WDs.
They explored the effect of the C/O ratio on the nucleosynthesis by
simulating both ${\rm C/O} = 50/50$ and $30/70$, which is more
representative of the ratio expected in actual WDs.  They considered
masses from $0.8$ to $1.1~M_{\sun}$ and metallicities from $0$ to
$2Z_{\sun}$.

\citet[][\citetalias{bra19}]{bra19} also conducted 1D simulations of detonations that
began at the centers of sub-$M_{\rm Ch}$ WDs.\footnote{\citetalias{bra19} also
  simulated DDT explosions, but we did not consider them because we
  only considered multi-dimensional simulations of deflagrations.}
They treated composition and metallicity in the same manner as \citetalias{leu19}.
They also explored the effect of reducing the reaction rate of
${}^{12}{\rm C} + {}^{16}{\rm O}$ by a factor of 10.  The models with
the reduced reaction rate are represented by $\xi_{\rm CO} = 0.9$ in
Table~\ref{tab:yields_theory}.  The models with the ``standard''
reaction rate have $\xi_{\rm CO} = 0.0$.

\vspace{4mm}

We first discuss differences between the models before comparing the
observations to theoretical predictions.  First, the metallicity of
the WD influences the yields, sometimes by a large amount.
\citet{tim03} studied how the ${}^{22}$Ne content, which depends
directly on initial metallicity, affects neutronization in the WD
core.  \citet{pir08} and \citet{cha08} further studied pre-explosion
``simmering,'' or convective burning in the core prior to explosion,
in near-$M_{\rm Ch}$ SN~Ia progenitors.  The effect of simmering is to
increase the neutron excess.  In effect, it makes the initial
metallicity of a $M_{\rm Ch}$ SN~Ia irrelevant below a threshold
metallicity.  The threshold metallicity imposed by simmering has
variously been estimated to be $2/3~Z_{\sun}$ \citep{pir08} or
$1/3~Z_{\sun}$ \citep{mar16}.  Both of these values are much larger
than the most metal-rich stars in our sample.  The $M_{\rm Ch}$ DDT
models computed at sub-solar metallicity in Table~\ref{tab:models} do
not include simmering.  Therefore, it is better to compare our
observationally inferred yields to the solar-metallicity DDT models.
From here on, we disregard the DDT models at sub-solar metallicity.
The pure deflagration models are available only at solar metallicity,
and the sub-$M_{\rm Ch}$ models are not subject to simmering.

Where possible, we show sub-$M_{\rm Ch}$ models interpolated to a
metallicity of $Z = 10^{-1.5}~Z_{\sun}$ so that it is easier to
compare with the observationally inferred yields
(Section~\ref{sec:Iacompare}), which we tabulate at ${\rm [Fe/H]} =
-1.5$.  However, some models are available only at solar metallicity.
The second column in Table~\ref{tab:yields_theory} specifies the
metallicity at which the theoretical yields are given.
Figure~\ref{fig:Iacompare} represents solar-metallicity models in
shades of red.  Models interpolated to a metallicity of $Z =
10^{-1.5}~Z_{\sun}$ are shown in shades of blue.

The effect of metallicity is especially apparent in the [Co/Fe] and
[Ni/Fe] sub-$M_{\rm Ch}$ yields of \citetalias{leu19}.  The yields of
the $1.10~M_{\sun}$ model reflect a metallicity of $Z =
10^{-1.5}~Z_{\sun}$, but modeled WDs of higher and lower masses have
solar metallicity.  The solar-metallicity models show a smooth
gradient in yields, but the low-metallicity model is offset to lower
abundance ratios.  This offset results from the neutron enhancement of
higher-metallicity models.  The extra neutrons allow the creation of
neutron-rich species, like ${}^{58}{\rm Ni}$.

The choice of ignition parameters also influences the yields
\citep[see the discussion by][]{fis15}.  In $M_{\rm Ch}$ models, a
larger number of ignition sites translates to greater pre-expansion of
the WD\@.  The resulting lower density leads to synthesis of more
intermediate-mass elements, like Si, and fewer Fe-group isotopes, such
as ${}^{56}$Ni, which decays to ${}^{56}$Fe, and ${}^{58}$Ni, which is
stable.  Therefore, a greater number of ignition sites---especially
off-center ignitions---leads to higher [Si/Fe] yields and lower
[Ni/Fe] yields.

The most notable feature that distinguishes the yields of different
classes of SNe~Ia is the difference in the [Ni/Fe] ratio between
near-$M_{\rm Ch}$ and sub-$M_{\rm Ch}$ explosions.  Higher-mass WDs
have higher densities, which permit more complete burning in nuclear
statistical equilibrium, including electron captures that increase the
neutron density.  A high neutron density favors the creation of
Fe-peak elements over intermediate-mass elements.  Even within the
sub-$M_{\rm Ch}$ simulations, WDs of higher mass (represented by
darker shades in Figure~\ref{fig:Iacompare}) produce lower [Si/Fe],
[Ca/Fe], and [Cr/Fe] ratios and higher [Co/Fe] and [Ni/Fe] ratios.
The only simulations that predict ${\rm [Ni/Fe]} < -0.01$ are
sub-$M_{\rm Ch}$ models.  Furthermore, all of the metal-poor
sub-$M_{\rm Ch}$ simulations predict ${\rm [Ni/Fe]} \le -0.30$.

\subsection{Comparison of Observationally Inferred Yields to
  Theoretical Yields}
\label{sec:Iacompare}

The preceding discussion suggests that ${\rm [Ni/Fe]}_{\rm Ia}$ is a
potential indicator of whether SNe~Ia result from the explosions of
near-$M_{\rm Ch}$ or sub-$M_{\rm Ch}$ WDs.  The observationally
inferred yields for different dSphs range from ${\rm [Ni/Fe]}_{\rm Ia}
= \nifeiamin_{-\nifeiaminerrlo}^{+\nifeiaminerrhi}$ (\nifeiamindsph)
to $\nifeiamax_{-\nifeiamaxerrlo}^{+\nifeiamaxerrhi}$
(\nifeiamaxdsph).  The inferred [Ni/Fe] yields are best explained by
sub-$M_{\rm Ch}$ detonations.  The [Ni/Fe] yields are not compatible
with $M_{\rm Ch}$ DDTs or pure deflagarations.  Under the assumption
that Type~Iax SNe are pure deflagrations, our results indicate that
Type~Iax SNe are not a major contributor to galactic chemical
evolution.  Other element ratios do not distinguish between
near-$M_{\rm Ch}$ or sub-$M_{\rm Ch}$ SNe~Ia as well as [Ni/Fe].
Nonetheless, the inferred yields for all elements and for all dSphs
are compatible with sub-$M_{\rm Ch}$ explosions.

As discussed in Section~\ref{sec:theoryyields}, the WD mass influences
the yield ratios.  The observationally inferred ${\rm [Ni/Fe]}_{\rm
  Ia}$ yields best match some of the more massive sub-$M_{\rm Ch}$
WDs, specifically the \citetalias{she18} and \citetalias{leu19} models
with masses of 1.00--1.10~$M_{\sun}$.  \citet{mcw18} pointed out that
[Si/Fe] is a strong indicator of WD mass.  Our estimates of ${\rm
  [Si/Fe]}_{\rm Ia}$ match the \citetalias{she18} and
\citetalias{bra19} models of WDs from $1.00~M_{\sun}$ to slightly
larger than $1.15~M_{\sun}$.  WDs of this mass are heavily outnumbered
by lower-mass WDs in old populations.  However, the inferred yields
reflect the stellar population of the dSphs at the time of star
formation, not the present time.  With the exception of Leo~II, the
galaxies in this study formed the majority of their stars in 1--2~Gyr
\citep{kir11b,wei14}.  The lifetime of a 1.7~$M_{\sun}$ star is about
1~Gyr \citep{pad93}, and such a star would make a WD of
0.59~$M_{\sun}$ \citep{kal08}.  Therefore, there would be plenty of
low-mass WDs during the active lifetime of the dSphs.  However, the
WDs must be given time to explode in order to relate this discussion
to SNe~Ia.  \citet{she17} modeled SNe~Ia as prompt double detonations
of WD binaries.  In that scenario, 90\% of SNe~Ia arise from WDs of
$>1~M_{\sun}$ in stellar populations with ages ranging from 0.3 to
1.0~Gyr.  The percentage drops to 50\% in stellar populations with
ages in the range 1--3~Gyr, which is the time frame in which many
dSphs formed most of their stars \citep{wei14}.

\begin{figure*}
\centering
\includegraphics[width=0.51\linewidth]{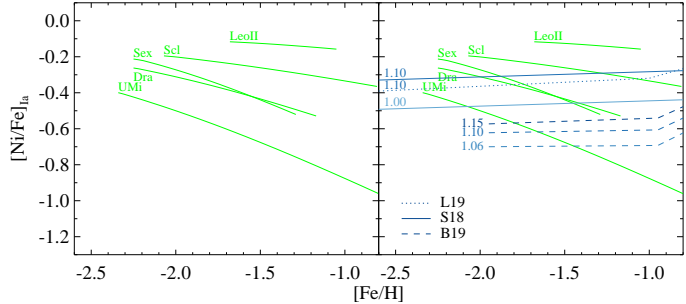}
\caption{Metallicity dependence of observationally inferred (green)
  and theoretical (shades of blue) [Ni/Fe] yields for sub-$M_{\rm Ch}$
  SNe~Ia.  The blue numbers indicate the WD mass in $M_{\sun}$.  Both
  panels show the same data, but the theoretical predictions are
  suppressed in the left panel for clarity.  The line style for each
  theoretical prediction reflects the study from which it came, and
  the shading indicates the initial WD mass, consistent with
  Figure~\ref{fig:Iacompare}.\label{fig:nife_metallicity}}
\end{figure*}

The WD metallicity also has an effect on yield ratios (also discussed
in Section~\ref{sec:theoryyields}).  We inferred SN~Ia yields as a
function of metallicity.  Figure~\ref{fig:nife_metallicity} compares
our inferences to the theoretical predictions.  In general, [Ni/Fe] is
predicted to increase with [Fe/H], but we inferred that the yield
decreases with metallicity.  However, the predicted metallicity
dependence at ${\rm [Fe/H]} < -1$ is very weak for WDs with masses
greater than $1~M_{\sun}$.  In fact, we might expect ${\rm
  [Ni/Fe]}_{\rm Ia}$ to decrease with metallicity because the average
mass of the exploding WDs decreases over time \citep{she17}.  As a
result, ${\rm [Ni/Fe]}_{\rm Ia}$ could reflect the yields of
lower-mass WDs as the galaxy ages and becomes more metal-rich.  This
hypothesis could be tested by convolving the exploding WD mass
function expected for a given age with the predicted SN~Ia yields.  We
save this analysis for the future.

\citetalias{she18} and \citetalias{bra19} considered two variables
that affected the SN~Ia yields.  \citetalias{she18} used two different
C/O ratios, and \citetalias{bra19} reduced the ${}^{12}{\rm C} +
{}^{16}{\rm O}$ reaction rate by a factor of 10.  (\citetalias{she18}
also experimented with the reaction rate, but we present only their
results using the ``standard'' rate.)  In most cases, these choices
result in differences of Fe-peak abundance ratios on the order of
$\lesssim 0.1$~dex.  In fact, \citetalias{bra19} specifically pointed
out that abundance ratios are less sensitive than absolute abundances
to the reaction rate.  Therefore, the C/O ratio and the ${\rm C}+{\rm
  O}$ reaction rate do not affect our conclusion about the dominant
role of sub-$M_{\rm Ch}$ SNe~Ia in dSphs.

Our conclusions are valid only for SNe~Ia that occurred during the
time in which the dSphs were forming stars.  The dSphs in our sample
formed the middle two thirds of their stars in durations ranging from
1~Gyr in Sculptor to 5~Gyr in Leo~II \citep{wei14}.  Some population
synthesis models predict that single-degenerate, $M_{\rm Ch}$ SNe~Ia
can be delayed by 0.6--1.9~Gyr \citep{rui09,rui11,yun10}.  However,
all of these long-delayed models fall short of explaining the
cosmological rate of SNe~Ia by at least two orders of magnitudes
\citep{nel13}.  Therefore, our measurements are likely sensitive to
the ``standard'' models of SNe~Ia that have shorter delays
\citep[i.e., those reviewed by][]{mao14}.  Even those models fall
short of the observed rate of SNe~Ia, but the discrepancy is less than
for the long-delayed models.  Regardless, it is important to note that
our measurements are not sensitive to SNe~Ia that are delayed by more
than the peak of star formation in the dSphs.  One example of such a
model is one that takes into account the effects of low metallicity on
the accretion rate and winds from accreting WDs \citep{kob09}.  We
discuss this model further at the end of Section~\ref{sec:discussion}.


\section{Discussion}
\label{sec:discussion}

We have used a galactic archaeological approach to inferring the
explosion mechanism of SNe~Ia.  We concluded that the
abundances---specifically [Ni/Fe]---in ancient dSphs are most
consistent with sub-$M_{\rm Ch}$ explosions.  Our approach assumed
that the galaxies were well-mixed at all times and that the IMF did
not change.  We made no assumptions about the form of the IMF\@.  We
also did not presuppose any theoretical SN yield for any abundance
ratio except for [Mg/Fe].

Ours is not the first galactic archaeological study to address the
nature of SNe~Ia.  Most of the previous studies focused on Mn because
it is particularly sensitive to the mass and metallicity of the
exploding WD \citep{sei13b}.  In fact, we will present our own Mn
abundance measurements in the near future.  \citet{nor12} measured Mn
abundances in four dSphs.  They measured sub-solar abundances of
[Mn/Fe] in all four dSphs, and they also found a trend with [Fe/H]\@.
They concluded that metallicity-dependent yields are the best
explanation for their observations.  However, they did not have the
benefit of the sizable menu of theoretical yields available today.  As
a result, they did not consider the mass of the exploding WD\@.

\citet{per10} discovered a new kind of explosion they called ``Ca-rich
SNe'' because of their high Ca/O nebular line ratios.  Their existence
in the luminosity gap between novae and SNe led \citet{kas12} to call
them ``Ca-rich gap transients.'' Our observations show that [Ca/Fe]
declines more slowly than [Si/Fe] in dwarf galaxies.  As a result, we
also find that ${\rm [Ca/Fe]}_{\rm Ia}$ is significantly higher than
${\rm [Si/Fe]}_{\rm Ia}$.  One interpretation is that SNe~Ia produce a
higher ratio of [Ca/Fe] than [Si/Fe].  In fact, this is the behavior
predicted by the sub-$M_{\rm Ch}$ models of \citetalias{she18} and
\citetalias{bra19}.  An alternative explanation is that there is a
second class of delayed explosions, i.e., Ca-rich transients.  While
our results do not require an additional nucleosynthetic source beyond
CCSNe and SNe~Ia, we cannot rule out that Ca-rich gap transients
contribute to the chemical evolution of dSphs.

So far, we have discounted Type~Iax SNe.  \citet{kob15} and
\citet{ces17} both explored the effect of Type Iax SNe on the chemical
evolution of dSphs.  They used measurements of $\alpha$ and Fe-peak
elements from \citet{ven12}, \citet{nor12}, \citet{ura15}, and others.
With the benefit of a one-zone chemical evolution model,
\citeauthor{kob15}\ concluded that a mixture of Type Iax and
sub-$M_{\rm Ch}$ explosions best explains the evolution of [Mn/Fe].
\citeauthor{ces17}\ abandoned the assumption of well-mixed gas.
Instead, they introduced stochasticity into their models as a way to
approximate the spread of [Mn/Fe] at a given [Fe/H]\@.  They
interpreted the observations of Ursa Minor as having a spread
consistent with their model, which invokes multiple sub-classes of
SN~Ia\@.

On the other hand, \citet{mcw18} presented a different interpretation
of the evolution of Fe-peak elements, including [Ni/Fe], in Ursa Minor
at fixed [Fe/H]\@.  They relied on \citeposs{coh10}\ high-resolution
spectroscopic measurements of 10 stars.  In contrast to \citet{ces17},
they did not find evidence for a spread in [Mn/Fe].  However, COS~171,
the star with the highest metallicity in the sample, has very low
[Mn/Fe] and [Ni/Fe] ratios.  The unusual abundances of this star are
reminiscent of Car~612 in the Carina dSph \citep{ven12}.
\citeauthor{mcw18}\ concluded that COS~171 was the result of the
explosion of a $0.95~M_{\sun}$ WD\@.  This interpretation for the
abundances of one star matches our interpretation for the abundances
of the ensemble of stars in several dSphs.  In other words, COS~171 is
not the result of a rare type of SN~Ia\@.  Instead, sub-$M_{\rm Ch}$
explosions are typical in dSphs.

Our conclusion is somewhat compatible with abundances of cold gas at
high redshift.  \citet{lu96} and \citet{coo15} measured Ni abundances
in damped Ly$\alpha$ (DLA) systems to be in the range $-0.4 \lesssim
{\rm [Ni/Fe]} \lesssim -0.1$.\footnote{However, \citet{pro02} measured
  $-0.2 \lesssim {\rm [Ni/Fe]} \lesssim +0.2$.  The Ni absorption
  lines measured in DLAs are very weak, which leads to large
  uncertainties in the [Ni/Fe] ratios.  Furthermore, both \citet{lu96}
  and \citet{pro02} mentioned that the oscillator strengths of the Ni
  transitions are not well known.}  The DLAs studied were all
metal-poor ($-2.4 \le {\rm [Fe/H]} \le -0.8$).  The agreement is
notable because we used a galactic archaeological technique at $z=0$,
but the DLA study was conducted at redshifts as high as $z=3$.  These
abundances possibly suggest that nucleosynthesis and chemical
evolution are similar across small, metal-poor galaxies.

\begin{figure}
\centering
\includegraphics[width=0.8\linewidth]{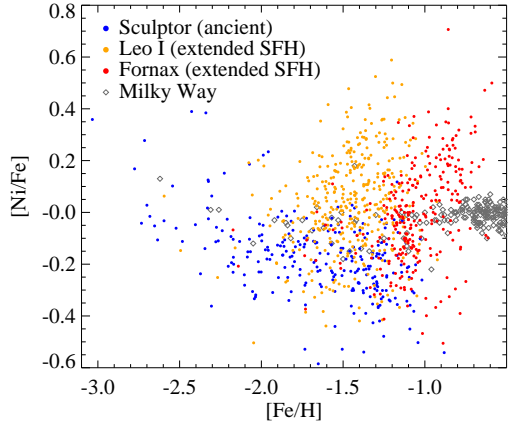}
\caption{Evolution of [Ni/Fe] with [Fe/H] in three dSphs.  Only stars
  with uncertainties less than 0.2~dex are shown.  The [Ni/Fe]
  evolution in the Milky Way \citep{ben14} is shown in gray
  diamonds.\label{fig:nife_feh}}
\end{figure}

Our findings regarding dwarf galaxies may not be extensible to more
massive galaxies, like the MW\@.  The [Ni/Fe] trend in the MW is
somewhat complex \citep[e.g.,][]{ven04,ben14,has17,lom19}.  In the
low-metallicity (${\rm [Fe/H]} \lesssim -1$) halo, the average value
of [Ni/Fe] is about $-0.1$.  At ${\rm [Fe/H]} > -1$, the dispersion in
[Ni/Fe] suddenly decreases, and the mean value is 0.0 until ${\rm
  [Fe/H]} \approx 0.0$.  (See Figure~\ref{fig:nife_feh}.)  At
super-solar metallicities, [Ni/Fe] smoothly increases with [Fe/H]\@.
One interpretation of this behavior is that the MW halo is composed of
dSphs with chemical evolution similar to Sagittarius
\citep[e.g.,][]{has17} and those studied in this work.  As a result,
their nucleosynthesis matches our interpretation, including the
dominance of sub-$M_{\rm Ch}$ SNe~Ia.  However, the MW disk has a
different type of stellar population.  In fact, \citet{sei13b}
previously argued that [Mn/Fe] abundances in the MW disk are too high
to be explained by sub-$M_{\rm Ch}$ SNe~Ia.  Perhaps the SNe~Ia that
exploded in the MW disk were of a different class than those that
exploded in the halo.

\citet{con14} found that giant elliptical galaxies with lower velocity
dispersion (i.e., those with lower [$\alpha$/Fe]) have lower
integrated-light abundances of [Ni/Fe].  This finding corroborates the
idea that galaxies with more SNe~Ia have lower [Ni/Fe] ratios.  One
interpretation of this result is that SNe~Ia are exploding below
$M_{\rm Ch}$ even in giant elliptical galaxies. However, the [Ni/Fe]
ratios found in those galaxies are much larger than those found in
this work for dwarf galaxies.  Furthermore, \citeauthor{con14}\ found
a weaker decrease in [Mn/Fe] with decreasing velocity dispersion.
Therefore, the interpretation of the integrated-light Fe-peak
abundances for giant galaxies is not as straightforward as for the
dwarf galaxies.

At even higher mass scales, Fe-peak abundances may be measured in the
intracluster gas in galaxy clusters with X-ray
spectroscopy. \citet{mer16a,mer16b} measured the abundances of
elements including Mn, Fe, and Ni in dozens of galaxy clusters.  They
concluded that the super-solar values of [Mn/Fe] and [Ni/Fe] precluded
a significant contribution of sub-$M_{\rm Ch}$ SNe~Ia.  Likewise, the
\citet{hit17} measured the abundances of Fe-peak elements from X-ray
spectra of the hot intracluster gas in the Perseus galaxy cluster.
They found solar ratios of [Mn/Fe] and [Ni/Fe].  Similar to
\citet{mer16b}, they concluded that the dominant source of Fe-peak
elements in the cluster is either near-$M_{\rm Ch}$ SNe~Ia or a
roughly equal mixture of near-$M_{\rm Ch}$ and sub-$M_{\rm Ch}$
SNe~Ia.  Bulk outflows from galaxies enriched the hot intracluster gas
whereas stars formed from the cold interstellar gas that did not
escape the galaxy.  Therefore, it may not be advisable to compare
galaxy cluster gas abundances to stellar abundances.  Regardless, the
X-ray measurements from galaxy clusters potentially corroborate a
scenario in which the dominant mode of SNe~Ia transitions from
sub-$M_{\rm Ch}$ at low galaxy mass (or low metallicity) to
near-$M_{\rm Ch}$ at higher galaxy mass.

\citet{kob98,kob15} and \citet{kob09} predicted that the rates and
explosion mechanisms of SNe~Ia depend on metallicity.  Specifically,
winds from WDs are more intense at higher metallicity.  Winds allow
the mass accretion rate to proceed slowly enough to be stable
\citep{hac96}.  Therefore, metallicities of ${\rm [Fe/H]} \gtrsim
-1.1$ could permit a single-degenerate WD to grow to $M_{\rm Ch}$ via
mass accretion from a red giant companion.  It is possible that the
dominant SN~Ia mechanism transitions from sub-$M_{\rm Ch}$ (or even
Type Iax, as suggested by \citealt{kob15}) at low metallicities to
near-$M_{\rm Ch}$ at high metallicities.  The observations of Fe-peak
abundances in some dwarf galaxies and in large galaxies, like the MW,
are consistent with this paradigm.

An alternative hypothesis is that the dominant class of SN~Ia depends
not on metallicity but on SFH\@.  Figure~\ref{fig:nife_feh} shows the
[Ni/Fe] evolution in Sculptor, Leo~I, and Fornax.  We have not
included Leo~I or Fornax in our discussion until now because they do
not conform well to our simple bi-linear model of chemical evolution
(Section~\ref{sec:model}).  The average metallicities of Leo~I and
Fornax are higher than that of Sculptor, but their metallicity ranges
overlap.  It is very interesting that Leo~I has higher values of
[Ni/Fe] than Sculptor {\it at the same metallicity}.  It seems that
metallicity is not the only variable that controls [Ni/Fe].  Even more
importantly, abundance ratios above ${\rm [Ni/Fe]} \sim +0.1$ cannot
be explained by any of the sub-$M_{\rm Ch}$ models listed in
Table~\ref{tab:yields_theory}.

\citet{wei14} measured the SFH of Sculptor to be exclusively ancient.
The SFHs of Leo~I and Fornax are extended.  Both formed stars steadily
for nearly the entire age of the universe.  Even so, both galaxies
show decreasing [$\alpha$/Fe] with increasing [Fe/H]
\citep{let10,kir11b,lem14}, which is evidence for increasing
enrichment by SNe~Ia.  Therefore, the enhanced, rising trends of
[Ni/Fe] for Leo~I and Fornax cannot be explained by a high ratio of
CCSNe to SNe~Ia.  Instead, it could be that a different, more delayed
class of SN~Ia dominates the [Ni/Fe] evolution in Leo~I and Fornax,
whereas the SNe~Ia in Sculptor are more prompt.  \citet{kob09}
presented a scenario where single-degenerate, $M_{\rm Ch}$ SNe~Ia
(i.e., those that produce ${\rm [Ni/Fe]} > 0$) explode later than
double-degenerate SNe~Ia (i.e., those that produce ${\rm [Ni/Fe]} <
0$).  It is possible that the [Ni/Fe] evolution in Leo~I and Fornax is
evidence of a transition from double-degenerate, sub-$M_{\rm Ch}$ to
single-degenerate, near-$M_{\rm Ch}$ SNe~Ia.


\section{Summary}
\label{sec:summary}

We have used our previously published catalogs of $\alpha$ and Fe-peak
abundances in dSphs \citep{kir10,kir18} to infer properties of SNe~Ia
in dwarf galaxies.  We used the evolution of [Mg/Fe] to quantify the
fraction ($f_{\rm Ia}$) of Fe produced in SNe~Ia.  We applied this
ratio to the observed evolution of other element ratios, including
[Ni/Fe], to infer the yields of SNe~Ia, e.g., ${\rm [Ni/Fe]}_{\rm
  Ia}$.  In the dSphs with short SFHs, the observationally inferred
yields of ${\rm [Si/Fe]}_{\rm Ia}$, ${\rm [Ca/Fe]}_{\rm Ia}$, ${\rm
  [Co/Fe]}_{\rm Ia}$, and ${\rm [Ni/Fe]}_{\rm Ia}$ are sub-solar,
whereas ${\rm [Cr/Fe]}_{\rm Ia}$ is approximately solar.  These yields
are consistent with the detonations of $\sim 1.0~M_{\sun}$ WDs.
However, the ${\rm [Ni/Fe]}_{\rm Ia}$ yields are inconsistent with the
explosions of near-$M_{\rm Ch}$ WDs.  Two dSphs with more extended
SFHs (Fornax and Leo~I) contain stars with enhanced [Ni/Fe] that
cannot be explained by sub-$M_{\rm Ch}$ SNe~Ia.

We concluded that the dominant explosion mechanism of SNe~Ia that
occurred before the end of star formation in short-lived dwarf
galaxies is the detonation of a sub-$M_{\rm Ch}$ WD\@.  This
conclusion cannot necessarily be generalized to the total population
of SNe~Ia.  Specifically, the evolution of [Ni/Fe] in the MW disk and
in dSphs with extended SFHs is different from that in ancient dSphs.

We did not observe a strong evolution of SN~Ia yields with
metallicity, but theoretical models do not predict a strong
metallicity dependence at ${\rm [Fe/H]} < -1$, which includes the vast
majority of stars in our sample.  The weak metallicity dependence we
did observe might instead confirm the prediction that lower-mass,
sub-$M_{\rm Ch}$ WDs explode at later times \citep{she17}.

We also observed a marked difference in the [Ni/Fe] evolution between
galaxies with different SFHs.  Specifically, [Ni/Fe] is sub-solar in
ancient galaxies, but [Ni/Fe] is higher and increases with [Fe/H] in
galaxies that formed stars for several Gyr.  The differences could
reflect a shift in the dominant type of SN~Ia (double-degenerate to
single-degenerate and sub-$M_{\rm Ch}$ to $M_{\rm Ch}$) for galaxies
with more extended SFHs.

In the future, we will measure Mn, which is perhaps the element most
sensitive to the mass and metallicity of the exploding WD\@.  The
evolution of [Mn/Fe] will be a powerful check on our conclusions,
which were based mostly on [Ni/Fe].  Our dataset also has several
other possible uses, such as constructing models of chemical evolution
that account for a range of WD masses (see
Section~\ref{sec:Iacompare}).  The data might also be useful in
deducing the SN~Ia delay time distribution.

\acknowledgments We thank Marten van Kerkwijk, Julianne Dalcanton, Ivo
Seitenzahl, Robert Fisher, Ken'ichi Nomoto, Mansi Kasliwal, Alex Ji,
and Norbert Werner for insightful conversations.  We also thank the
anonymous referee for a helpful report.  This material is based upon
work supported by the National Science Foundation under grant
No.\ AST-1614081.  E.N.K. gratefully acknowledges support from a
Cottrell Scholar award administered by the Research Corporation for
Science Advancement as well as funding from generous donors to the
California Institute of Technology.

We are grateful to the many people who have worked to make the Keck
Telescope and its instruments a reality and to operate and maintain
the Keck Observatory.  The authors wish to extend special thanks to
those of Hawaiian ancestry on whose sacred mountain we are privileged
to be guests.  Without their generous hospitality, none of the
observations presented herein would have been possible.  We express
our deep gratitude to the staff at academic and telescope facilities
whose labor maintains spaces for scientific inquiry.

\facility{Keck:II (DEIMOS)}
\software{MOOG \citep{sne73,sne12}}

\bibliography{fepeak_Ia}
\bibliographystyle{apj}

\startlongtable
\begin{deluxetable*}{lccccccc}
\tablewidth{0pt}
\tablecolumns{8}
\tablecaption{Theoretically Predicted Yields\label{tab:yields_theory}}
\tablehead{\colhead{Model} & \colhead{$\log (Z/Z_{\sun})$} & \colhead{[Mg/Fe]} & \colhead{[Si/Fe]} & \colhead{[Ca/Fe]} & \colhead{[Cr/Fe]} & \colhead{[Co/Fe]} & \colhead{[Ni/Fe]}}
\startdata
\cutinhead{DDT(S13)}
N1                                                             & \phs 0.0 & $-2.16$ & $-0.99$ & $-1.16$ & $-0.57$ & $-0.87$ & $+0.06$ \\
N3                                                             & \phs 0.0 & $-1.87$ & $-0.81$ & $-0.98$ & $-0.49$ & $-0.77$ & $+0.06$ \\
N10                                                            & \phs 0.0 & $-1.81$ & $-0.59$ & $-0.76$ & $-0.32$ & $-0.66$ & $+0.12$ \\
N100                                                           & \phs 0.0 & $-1.38$ & $-0.14$ & $-0.40$ & $+0.02$ & $-0.57$ & $+0.25$ \\
N200                                                           & \phs 0.0 & $-0.92$ & $+0.04$ & $-0.42$ & $-0.02$ & $-0.45$ & $+0.39$ \\
N1600                                                          & \phs 0.0 & $-0.92$ & $+0.11$ & $-0.37$ & $-0.04$ & $-0.35$ & $+0.44$ \\
\cutinhead{def(F14)}
N1def                                                          & \phs 0.0 & $-1.30$ & $-0.61$ & $-0.94$ & $-0.27$ & $-0.59$ & $+0.32$ \\
N3def                                                          & \phs 0.0 & $-1.16$ & $-0.50$ & $-0.87$ & $-0.20$ & $-0.49$ & $+0.38$ \\
N10def                                                         & \phs 0.0 & $-1.25$ & $-0.54$ & $-0.89$ & $-0.19$ & $-0.52$ & $+0.37$ \\
N100def                                                        & \phs 0.0 & $-1.01$ & $-0.46$ & $-0.87$ & $-0.15$ & $-0.46$ & $+0.39$ \\
N200def                                                        & \phs 0.0 & $-1.09$ & $-0.49$ & $-0.90$ & $-0.15$ & $-0.42$ & $+0.43$ \\
N1600def                                                       & \phs 0.0 & $-1.11$ & $-0.52$ & $-0.93$ & $-0.15$ & $-0.33$ & $+0.46$ \\
\cutinhead{DDT(L18)}
WDD2                                                           & \phs 0.0 & $-1.73$ & $-0.24$ & $-0.17$ & $+0.26$ & $-1.48$ & $-0.01$ \\
DDT $1 \times 10^9$~g~cm$^{-3}$                                & \phs 0.0 & $-2.36$ & $-0.29$ & $-0.30$ & $-0.22$ & $-4.26$ & $+0.11$ \\
DDT $3 \times 10^9$~g~cm$^{-3}$                                & \phs 0.0 & $-2.59$ & $-0.29$ & $-0.38$ & $-0.10$ & $-1.62$ & $+0.20$ \\
DDT $5 \times 10^9$~g~cm$^{-3}$                                & \phs 0.0 & $-2.53$ & $-0.21$ & $-0.42$ & $+0.28$ & $-0.85$ & $+0.26$ \\
\cutinhead{def(L18)}
W7                                                             & \phs 0.0 & $-2.01$ & $-0.46$ & $-0.66$ & $-0.14$ & $-1.81$ & $+0.16$ \\
def $1 \times 10^9$~g~cm$^{-3}$                                & \phs 0.0 & $-1.59$ & $-0.63$ & $-0.93$ & $-0.54$ & $-4.09$ & $+0.25$ \\
def $3 \times 10^9$~g~cm$^{-3}$                                & \phs 0.0 & $-1.85$ & $-0.85$ & $-1.18$ & $-0.10$ & $-1.30$ & $+0.36$ \\
def $5 \times 10^9$~g~cm$^{-3}$                                & \phs 0.0 & $-2.08$ & $-1.02$ & $-1.32$ & $+0.32$ & $-0.73$ & $+0.29$ \\
\cutinhead{sub(L19)}
$0.90~M_{\sun}$, $M_{\rm He} = 0.15~M_{\sun}$                  & \phs 0.0 & $-0.30$ & $+0.38$ & $+0.15$ & $+0.86$ & $-0.21$ & $+0.09$ \\
$0.95~M_{\sun}$, $M_{\rm He} = 0.15~M_{\sun}$                  & \phs 0.0 & $-0.88$ & $-0.08$ & $-0.30$ & $+0.61$ & $-0.41$ & $+0.05$ \\
$1.00~M_{\sun}$, $M_{\rm He} = 0.10~M_{\sun}$                  & \phs 0.0 & $-0.88$ & $-0.09$ & $-0.33$ & $+0.53$ & $-0.39$ & $+0.03$ \\
$1.05~M_{\sun}$, $M_{\rm He} = 0.10~M_{\sun}$                  & \phs 0.0 & $-1.19$ & $-0.32$ & $-0.54$ & $+0.41$ & $-0.41$ & $+0.06$ \\
$1.10~M_{\sun}$, $M_{\rm He} = 0.10~M_{\sun}$                  &   $-1.5$ & $-1.32$ & $-0.48$ & $-0.51$ & $+0.31$ & $-0.83$ & $-0.34$ \\
$1.15~M_{\sun}$, $M_{\rm He} = 0.10~M_{\sun}$                  & \phs 0.0 & $-2.09$ & $-0.67$ & $-0.72$ & $+0.08$ & $-0.43$ & $+0.04$ \\
$1.20~M_{\sun}$, $M_{\rm He} = 0.05~M_{\sun}$                  & \phs 0.0 & $-1.93$ & $-0.70$ & $-0.79$ & $-0.12$ & $-0.52$ & $+0.12$ \\
\cutinhead{sub(S18)}
$0.85~M_{\sun}$, ${\rm C/O} = 50/50$                           &   $-1.5$ & $-0.82$ & $+0.61$ & $+0.61$ & $+0.58$ & $-2.04$ & $-1.70$ \\
$0.90~M_{\sun}$, ${\rm C/O} = 50/50$                           &   $-1.5$ & $-1.37$ & $+0.24$ & $+0.30$ & $+0.47$ & $-2.48$ & $-1.92$ \\
$1.00~M_{\sun}$, ${\rm C/O} = 50/50$                           &   $-1.5$ & $-2.19$ & $-0.23$ & $-0.11$ & $+0.12$ & $-0.70$ & $-0.46$ \\
$1.10~M_{\sun}$, ${\rm C/O} = 50/50$                           &   $-1.5$ & $-3.01$ & $-0.63$ & $-0.44$ & $-0.17$ & $-0.55$ & $-0.30$ \\
$0.85~M_{\sun}$, ${\rm C/O} = 30/70$                           &   $-1.5$ & $-0.46$ & $+0.75$ & $+0.79$ & $+0.64$ & $-1.51$ & $-1.55$ \\
$0.90~M_{\sun}$, ${\rm C/O} = 30/70$                           &   $-1.5$ & $-1.17$ & $+0.30$ & $+0.42$ & $+0.52$ & $-2.30$ & $-1.89$ \\
$1.00~M_{\sun}$, ${\rm C/O} = 30/70$                           &   $-1.5$ & $-2.14$ & $-0.20$ & $-0.02$ & $+0.19$ & $-0.80$ & $-0.54$ \\
$1.10~M_{\sun}$, ${\rm C/O} = 30/70$                           &   $-1.5$ & $-3.07$ & $-0.59$ & $-0.37$ & $-0.10$ & $-0.59$ & $-0.32$ \\
\cutinhead{sub(B19)}
$0.88~M_{\sun}$, $\xi_{\rm CO} = 0.9$                          &   $-1.5$ & $-0.52$ & $+0.42$ & $+0.66$ & $+0.48$ & $-2.77$ & $-1.85$ \\
$0.97~M_{\sun}$, $\xi_{\rm CO} = 0.9$                          &   $-1.5$ & $-1.29$ & $-0.09$ & $+0.25$ & $+0.30$ & $-1.55$ & $-1.59$ \\
$1.06~M_{\sun}$, $\xi_{\rm CO} = 0.9$                          &   $-1.5$ & $-1.94$ & $-0.46$ & $-0.07$ & $+0.02$ & $-0.82$ & $-0.70$ \\
$1.10~M_{\sun}$, $\xi_{\rm CO} = 0.9$                          &   $-1.5$ & $-2.24$ & $-0.62$ & $-0.20$ & $-0.11$ & $-0.74$ & $-0.62$ \\
$1.15~M_{\sun}$, $\xi_{\rm CO} = 0.9$                          &   $-1.5$ & $-2.64$ & $-0.83$ & $-0.39$ & $-0.27$ & $-0.67$ & $-0.56$ \\
$0.88~M_{\sun}$, $\xi_{\rm CO} = 0.0$                          &   $-1.5$ & $-0.38$ & $+0.60$ & $+0.64$ & $+0.48$ & $-2.68$ & $-1.82$ \\
$0.97~M_{\sun}$, $\xi_{\rm CO} = 0.0$                          &   $-1.5$ & $-1.22$ & $+0.02$ & $+0.18$ & $+0.28$ & $-1.52$ & $-1.57$ \\
$1.06~M_{\sun}$, $\xi_{\rm CO} = 0.0$                          &   $-1.5$ & $-1.86$ & $-0.36$ & $-0.15$ & $-0.01$ & $-0.81$ & $-0.68$ \\
$1.10~M_{\sun}$, $\xi_{\rm CO} = 0.0$                          &   $-1.5$ & $-2.16$ & $-0.53$ & $-0.28$ & $-0.13$ & $-0.75$ & $-0.61$ \\
$1.15~M_{\sun}$, $\xi_{\rm CO} = 0.0$                          &   $-1.5$ & $-2.56$ & $-0.75$ & $-0.46$ & $-0.29$ & $-0.73$ & $-0.60$ \\
\enddata
\end{deluxetable*}

\end{document}